\documentclass[aps,prev,preprintnumbers,floatset,nofootinbib,twocolumn]{revtex4-1}
\usepackage[english]{babel}
\usepackage{bm}
\usepackage{times}
\usepackage{ulem}
\usepackage{hyperref}
\hypersetup{
    colorlinks=true,
    linkcolor=blue,
    filecolor=magenta,      
    urlcolor=blue,
    citecolor=blue,
    pdftitle={Sharelatex Example},
    pdfpagemode=FullScreen
    }
\usepackage{listings}
\usepackage{slashed}
\usepackage{color}
\usepackage{appendix}
\usepackage{mathtools}
\usepackage{epsfig}
\usepackage{dcolumn}
\usepackage{textcomp}
\usepackage{appendix} 
\usepackage{multirow}
\usepackage{graphicx}
\usepackage{soul}
\usepackage{subfigure}
\usepackage{xcolor}

\newcommand{\fig}[1]{Fig.~\ref{#1}} 
 
\newcommand{\eq}[1]{Eq.~(\ref{#1})} 
 
\newcommand{\Deivid}[1]{{\color{black}#1}}

\begin{document} 

\title{The Hubble Constant Troubled by Dark Matter in Non-Standard Cosmologies}
\author{J. S. Alcaniz$^{1,\dagger}$}
\author{J. P. Neto$^{2,3,4,\ddagger}$}
\author{F. S. Queiroz$^{3,4,5,\S}$}
\author{D. R. da Silva$^{6,*}$}
\author{R. Silva$^{2,3,\P}$}

\affiliation{$^{1}$ Observatório Nacional, 20921-400, Rio de Janeiro, RJ, Brasil}
\affiliation{$^2$ Instituto de F\'isica Gleb Wataghin, Universidade Estadual de Campinas, 13083-859, Campinas, SP, Brasil}
\affiliation{$^3$ International Institute of Physics, Universidade Federal do Rio Grande do Norte,  59078-970, Natal, RN, Brasil}
\affiliation{$^4$ Departamento de F\'isica, Universidade Federal do Rio Grande do Norte, 59078-970, Natal, RN, Brasil}
\affiliation{$^5$ Millennium Institute for Subatomic Physics at High-Energy Frontier (SAPHIR), Fernandez Concha 700, Santiago, Chile}
\affiliation{$^6$ Departamento de Fisica, Universidade Federal da Paraiba, 58051-970, Jo\~ao Pessoa, PB, Brasil}

\affiliation{$^{\dagger}$ alcaniz@on.br}
\affiliation{$^{\ddagger}$ jpneto@ifi.unicamp.br}
\affiliation{$^{\S}$ farinaldo.queiroz@ufrn.br}
\affiliation{$^{*}$ deivid.rodrigo@academico.ufpb.br}
\affiliation{$^{\P}$ raimundosilva@fisica.ufrn.br}

\begin{abstract}
The Standard Cosmological Model has experienced tremendous success at reproducing observational data by assuming a universe dominated by a cosmological constant and dark matter in a flat geometry. However, several studies, based on local measurements, indicate that the universe is expanding too fast, in disagreement with the Cosmic Microwave Background. Taking into account combined data from CMB, Baryon Acoustic Oscillation, and type Ia Supernovae, we show that if the mechanism behind the production of dark matter particles has at least a small non-thermal origin, one can induce larger values of the Hubble rate $H_0$, within the $\Lambda$CDM, to alleviate the trouble with $H_0$. In the presence of non-standard cosmology, however, we can fully reconcile CMB and local measurements and reach $H_0=70-74\, {\rm km s^{-1} Mpc^{-1}}$.
\noindent

\end{abstract}


\maketitle
\flushbottom

\section{\label{In} Introduction}

The standard $\Lambda$CDM describes an accelerated expansion of the universe that is currently dominated by dark matter and a cosmological constant, and from small density perturbations powered by inflation explain the formation of structures in the universe. This simple scenario has experienced a great concordance with cosmological data \cite{Aghanim:2018eyx}. One of the pillars of the $\Lambda$CDM model is the Cosmic Microwave Background (CMB), which acquired unprecedented precision with the Planck mission \cite{Planck:2015fie}. The CMB stands for the photons from the early universe that traveled long distances after their decoupling from the thermal bath carry information from the early universe, but which is also impacted by late-time universe physics as they propagate to us. The CMB features a near perfect black-body spectrum. The information encoded in the CMB data from polarization, temperature, and lensing is typically interpreted in terms of a standard spatially-flat 6-parameter $\Lambda$CDM cosmology. Planck satellite data \cite{Planck:2018vyg} of the Cosmic Microwave Background (CMB) anisotropies, combined with Atacama Cosmology Telescope \cite{ACT:2020gnv}, and South Pole Telescope \cite{SPT-3G:2021wgf} observations have confirmed that the $\Lambda$CDM model offers the best description of the universe, but at the same time gave rise to hints of physics beyond the $\Lambda CDM$. The most statistically significant  anomaly relies in the Hubble constant $H_0$ \cite{DiValentino2021}. The Hubble constant $H_0$ is the present expansion rate defined as $H_0=H(z=0)$ with $H=a^{-1}\frac{da}{dt}$, where $a^{-1}=1+z$.

In other words, The Hubble rate problem concerns about the discrepancy between the Hubble rate inferred from the CMB data and the one obtained from local measurements. In particular, Planck collaboration fits the CMB data using a 6 parameters model based on the $\Lambda$CDM cosmology, and from this fit infered (model-dependent) Hubble constant to be $H_0= 67.27 \pm 0.6 \, km s^{-1} Mpc^{-1}$ \cite{Planck:2018vyg}, whereas local measurements favor larger values that range from  $H_0= 71.8 \, km s^{-1} Mpc^{-1}$ up to  $H_0= 77 \, km s^{-1} Mpc^{-1}$, depending on the dataset used \cite{Kenworthy:2022jdh}. We will adopt a more conservative value  $H_0 = 73.2 \pm 1.3 \, km s^{-1} Mpc^{-1}$ \cite{ref:Anchordoqui2021} as a reference.

Several proposals have been put forth concerning the Hubble rate problem \cite{Shah:2021onj,DiValentino:2021izs,DiValentino:2020zio}, but in the realm of particle physics they typically rely on new interactions involving the Standard Model (SM) neutrinos or decaying dark matter models \cite{Abdalla:2022yfr}. In this work, we take a different route, and introduce a non-thermal production mechanism of dark matter to increase the relativistic degrees of freedom and consequently raise $H_0$ \cite{Hooper:2011aj,Kelso:2013paa}.

The idea consist of invoking a non-thermal dark matter production via the decay $\chi^\prime \rightarrow \chi +\nu$, where $\chi$ is stable and reproduces the correct dark matter relic density indicated by Planck collaboration \cite{Planck:2018vyg}. We will assume that $m_{\chi^\prime} \gg m_{\chi}$, thus the dark matter particle will be relativistic at first but as the universe expands it cools and becomes a standard cold relic at the matter-radiation equality for structure formation purposes. If a large fraction of the overall dark matter abundance comes from the decay of $\chi^\prime$, the change in the matter power spectrum is sufficiently large, in disagreement with Lyman-$\alpha$ observations \cite{Allahverdi:2014bva}. This fact is also important to avoid conflict with structure formation \cite{Bringmann:2018jpr}.  We will assume throughout that just a fraction of the dark matter abundance stems from this mechanism. We will carry out study in a model independent way. Because a fraction of dark matter particles were relativistic, they will mimic the effect of extra dark radiation, i.e relativistic degrees of freedom, $N_{eff}$. As the Hubble constant inferred from CMB observations is positively correlated with $N_{eff}$, an increase in $N_{eff}$ translates into a larger $H_{0}$. 

In the past years, this relation between $H_0$ and $N_{eff}$ has been explored within the $\Lambda$CDM model. However, recent studies show that one cannot find sufficiently larger values of $H_0$ in agreement with local measurements via $N_{eff}$ \cite{Abdalla:2022yfr}. Physics beyond the $\Lambda$CDM is needed. Having that in mind, we use combined data from Planck, BAO and Supernovae IA observations to determine what is the region of parameter in which our mechanism can increase $H_0$ and reconcile CMB and local measurements. It will be clear later on, that $\chi$ cannot be any particle, it ought to be a cold dark matter particle that reproduces well the cosmological data. In this way, our solution to $H_0$ is tied to dark matter, conversely to hidden neutrino interactions. The neutrino appearing in the final state in the $\chi^\prime \rightarrow \chi +\nu$ decay is merely a choice, and it does not impact our overall conclusions. One could replace the neutrino by a photon or any other particle from the Standard Model of particle physics. 

This work is structured as follows: We start by reviewing theoretical aspects of the mechanism; later we show that without non-standard cosmology, one cannot find values of $H_0$ large than $70 {\rm kms^{-1}Mpc^{-1}}$; further we exhibit the region of parameter in which we can reconcile CMB and local measurements of $H_0$; lastly take into account Big Bang Nucleosynthesis and CMB constraints on energy injection episodes and draw our conclusions.

\section{\label{Mo} Dark matter particles as the source of dark radiation}

We show how this non-thermal dark matter production mechanism can source dark radiation and solve the $H_0$ problem.  We remind that the radiation density $(\rho_{rad})$ is determined by the photon's temperature $(T)$ and the relativistic degrees of freedom $(g_*)$, i.e.,
\begin{equation}
    \rho_{rad} = \frac{\pi^2}{30}g_*T^4.
    \label{eqrad}
\end{equation}

In a radiation-dominated universe phase where only photons and neutrinos are ultrarelativistic the relation between photons and neutrinos temperature is $(4/11)^{1/3}$. As photons have two polarization states, and neutrinos are only left-handed in the standard model (SM); therefore, we write $g_*$ in the following way,
\begin{equation}
    g_* = 2 + \frac{7}{4} \left(\frac{4}{11} \right)^{4/3}N_{eff}.
    \label{eqNeff}
\end{equation}where $N_{eff}$ is the effective number of relativistic neutrino species, where in the $\Lambda$CDM is $N_{eff}=3$. 

In a more general setting there could be new light species contributing to $N_{eff}$, or some new physics interactions with neutrinos that will alter the neutrino decoupling temperature, or as in our case, some particles mimicking the effects of neutrinos. As we are trying to raise $H_0$ by increasing $N_{eff}$, $\Delta N_{eff}$ tell us how much extra radiation we are adding to the universe via our mechanism. In other words,

\begin{equation}
    \Delta N_{eff} = \frac{\rho_{extra}}{\rho_{1\nu}}.
    \label{eq:deltaN_general}
\end{equation}where ${\rho_{1\nu}}$ is the radiation density generated by an extra neutrino species.

Hence, in principle, we may reproduce the effect of an extra neutrino species by adding any other kind of radiation source. Calculating the ratio between one neutrino species density and cold dark matter density at the matter-radiation equality $(t = t_{eq})$ we get,
\begin{equation}
   \left. \frac{\rho_{1\nu}}{\rho_{DM}} \right|_{t = t_{eq}} = \frac{\Omega_{\nu,0}\rho_c}{3a^4_{eq}} \times \left(\frac{\Omega_{DM,0}\rho_c}{a^3_{eq}}\right)^{-1} = 0.16.
    \label{eq:0.16}
\end{equation}where we used $\Omega_{\nu,0} = 3.65 \times 10^{-5}$, $\Omega_{DM,0} = 0.265$ and $a_{eq} = 3 \times 10^{-4}$ \cite{ref:Planck2018}.

The above equation tells us that one extra neutrino species represents $16\%$ of the dark matter density at the matter-radiation equality. Assuming $\chi$ is produced via two body decays of a mother particle $\chi'$, where $\chi' \rightarrow \chi + \nu$. In $\chi'$ resting frame, the 4-momentum of particles are,
\begin{gather*}
    p_{\chi'} = \left(m_{\chi'}, \bm{0} \right),\\
    p_{\chi} = \left(E(\bm{p}), \bm{p} \right),\\
    p_{\nu} = \left(\left| \bm{p} \right|, -\bm{p} \right).
\end{gather*}
Therefore, the 4-momentum conservation implies,
\begin{equation}
    E_{\chi}(\tau) = m_{\chi} \left( \frac{m_{\chi'} }{2m_{\chi}} + \frac{m_{\chi} }{2m_{\chi'}}  \right) \equiv m_{\chi}\gamma_{\chi}(\tau),
\end{equation}where $\tau$ is the lifetime of the mother particle $\chi'$. We highlight that we will adopt the instant decay approximation.

Using this result and the fact that the momentum of a particle is inversely proportional to the scale factor, we obtain,
\begin{equation*}
    \begin{split}
        &E^2_{\chi} - m^2_{\chi} = \bm{p}^2_{\chi} \propto \frac{1}{a^2}\\
        &\Rightarrow  \left( E^2_{\chi}(t) - m^2_{\chi} \right)a^2(t) = \left( E^2_{\chi}(\tau) - m^2_{\chi} \right)a^2(\tau)\\
        &\Rightarrow \frac{E_{\chi}(t)}{m_{\chi}} = \left[ 1 + \left( \frac{a(\tau)}{a(t)} \right)^2 \left( \gamma^2_{\chi}(\tau) - 1\right)\right]^{1/2} \equiv \gamma_{\chi}(t).
    \end{split}
\end{equation*}

We are considering that the universe is in radiation domination phase, where $a(\tau)/a(t) = \sqrt{\tau/t}$. In this way, the dark matter Lorentz factor becomes,
\begin{equation}
    \gamma_{\chi}(t) = \sqrt{\frac{ (m^2_{\chi} - m^2_{\chi'})^2 }{4m^2_{\chi}m^2_{\chi'}} \left( \frac{\tau}{t} \right) + 1 }.
\end{equation}

In the nonrelativistic regime, $m_{\chi}$ is the dominant contribution to the energy of a particle. Thus, rewriting the dark matter energy we find,
\begin{equation*}
    E_{\chi} = m_{\chi}\left( \gamma_{\chi} -1 \right) + m_{\chi}.
\end{equation*}

Hence, in the ultrarelativistic regime $m_{\chi}\left( \gamma_{\chi} -1 \right)$ dominates. Consequently, the total energy of the dark matter particle can be written as,
\begin{equation*}
    E_{DM} = N_{HDM}m_{\chi}\left( \gamma_{\chi} -1 \right) + N_{CDM}m_{\chi}.
\end{equation*}
Here, $N_{HDM}$ is the total number of relativistic dark matter particles (hot particles), whereas $N_{CDM}$ is the total number of nonrelativistic DM (cold particles). Obviously, $N_{HDM} \ll N_{CDM}$ to be consistent with the cosmological data. The ratio between relativistic and nonrelativistic dark matter density energy is,
\begin{equation}
    \frac{\rho_{HDM}}{\rho_{CDM}} = \frac{N_{HDM}m_{\chi}\left( \gamma_{\chi} -1 \right)}{N_{CDM}m_{\chi}} \equiv f\left( \gamma_{\chi} -1 \right).
    \label{eqrhoHDM}
\end{equation}

Consequently, $f$ is the fraction of dark matter particles which are produced via this non-thermal process. As aforementioned, $f$ ought to be small, but we do not have to assume a precise value for it, but it will be of the order of $0.01$. This fact will be clear further.

Using Eq.\eqref{eq:deltaN_general} and Eq.\eqref{eqrhoHDM}, we find that the extra radiation produced via this mechanism is,
\begin{equation}
    \Delta N_{eff} = \lim_{t \to t_{eq}} \frac{f\left( \gamma_{\chi} -1 \right)}{0.16},
    \label{EqNeffinicial}
\end{equation}where we used Eq. \eqref{eq:0.16} and we wrote $\rho_{CDM} = \rho_{\chi}$.

In the regime $m_{\chi'} \gg m_{\chi}$, we simplify,
\begin{equation*}
    \gamma_{\chi}(t_{eq}) -1 \approx \gamma_{\chi}(t_{eq}) \approx \frac{m_{\chi^\prime}}{2m_{\chi}} \sqrt{\frac{\tau}{t_{eq}}},
\end{equation*}and Eq.\eqref{EqNeffinicial} reduces to,
\begin{equation}
        \Delta N_{eff} \approx 2.5 \times 10^{-3}\sqrt{\frac{\tau}{10^{6}s}} \times f\frac{m_{\chi'}}{m_{\chi}}.
    \label{eq:deltaN}
\end{equation} with $t_{eq} \approx 50 000 ~ \text{years} \approx 1.6 \times 10^{12} ~s$.

From Eq.\eqref{eq:deltaN}, we conclude that the  $\Delta N_{eff} \sim 1$ implies in a larger ratio $f\, m_{\chi^\prime}/m_{\chi}$ for a decay lifetime $\tau \sim 10^4- 10^8s$. Notice that our overall results rely on two free parameters: i) the lifetime, $\tau$, and ii) $f\, m_{\chi^\prime}/m_{\chi}$.

\section{Relation between Hubble constant and Dark Radiation}

\subsection*{Case 1: Within the $\Lambda$CDM}

\begin{figure}[htb!]
    \centering
    \subfigure[]{
        \includegraphics[width=\columnwidth]{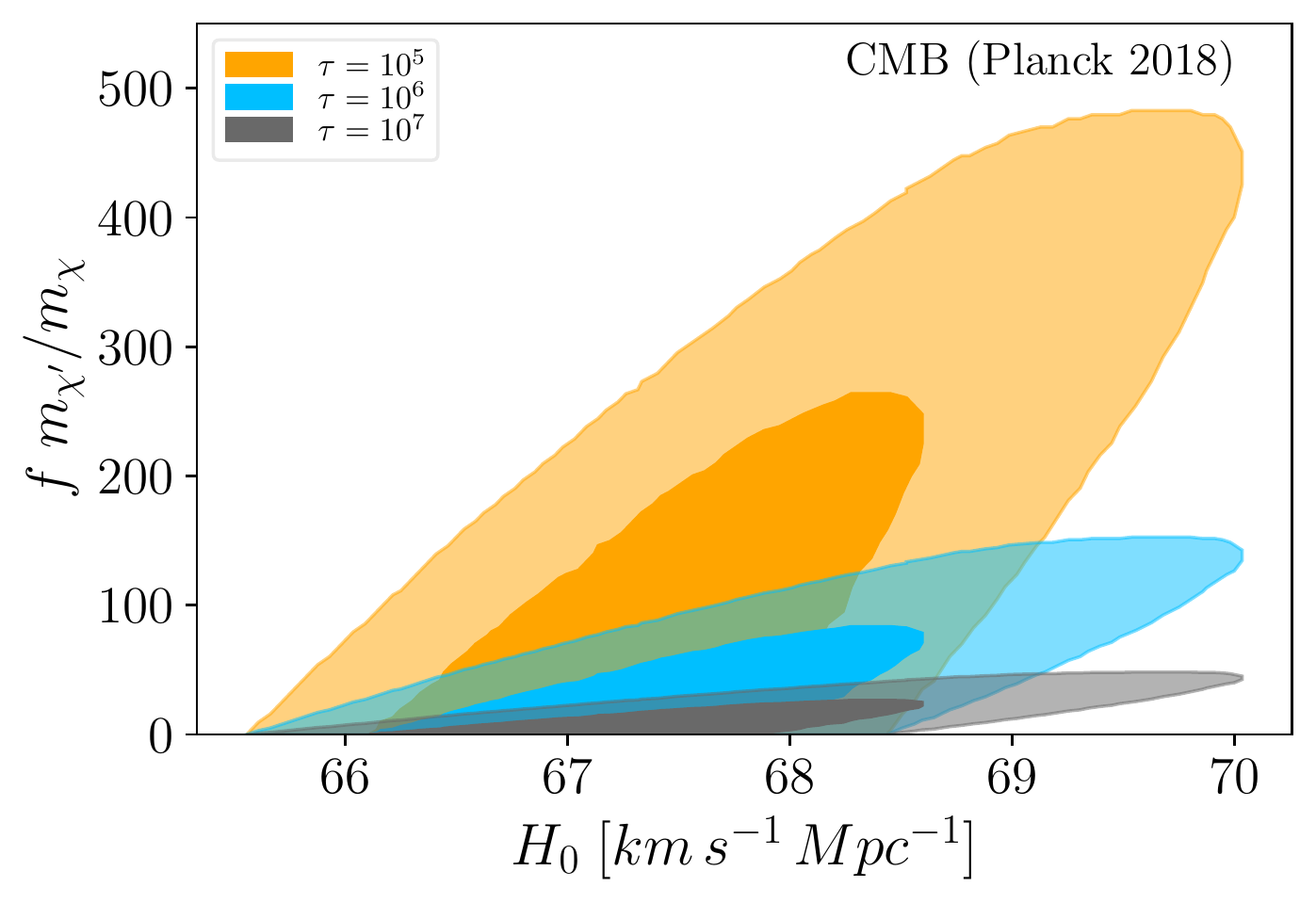}
        \label{fig:lambdaCDM_case_a}
    }
    \subfigure[]{
        \includegraphics[width=\columnwidth]{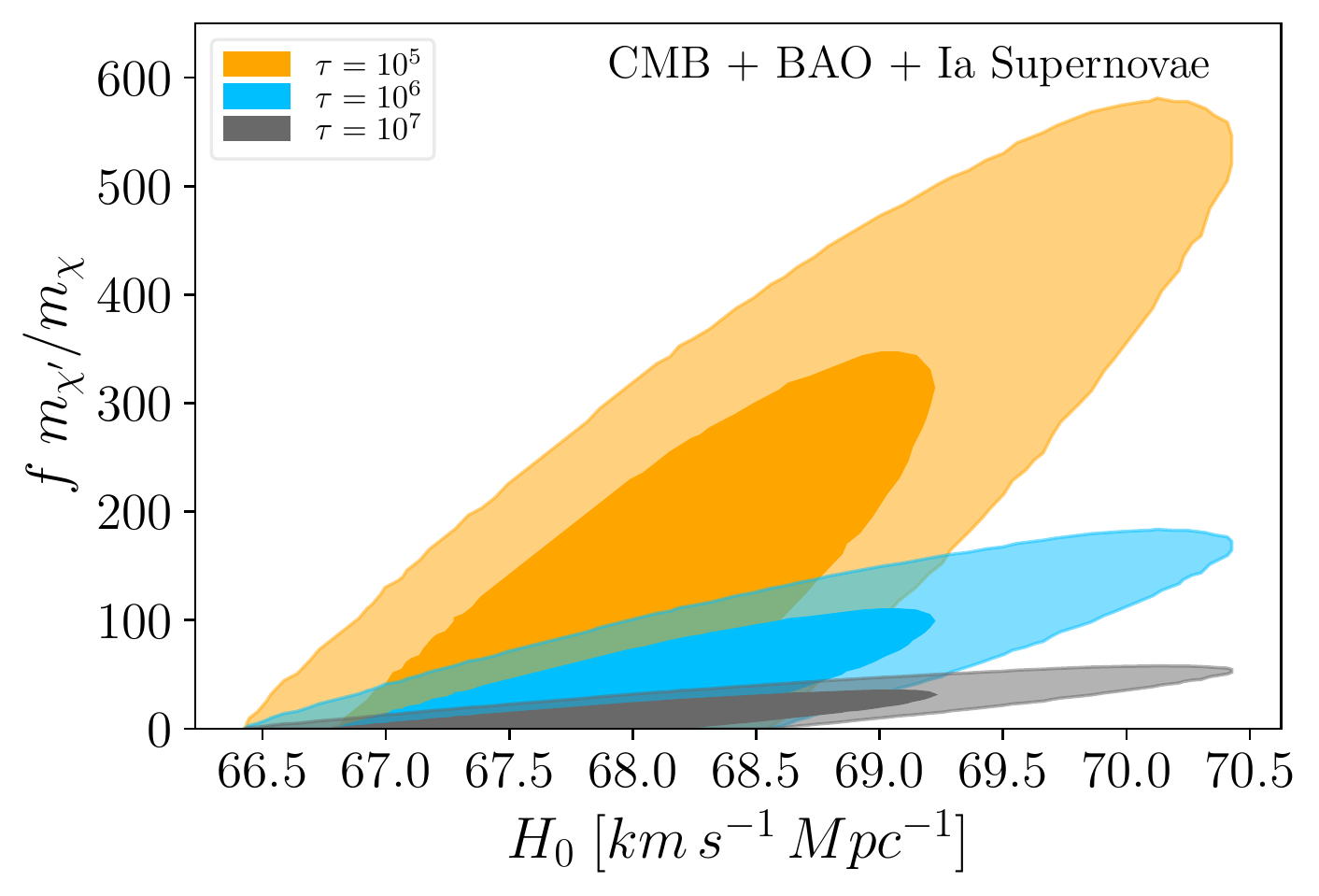}
        \label{fig:lambdaCDM_case_b}
    }
    
    \caption{Non-thermal production of a fraction, $f$, of dark matter particles via the $\chi^\prime \rightarrow \chi +\nu$ mechanism. Within the $\Lambda$CDM model we plot the region of parameter space in terms of $f m_{\chi^\prime}/m_\chi$ and $H_0$ for different decay lifetimes, either considering Planck data only \textbf{(a)} or combining it with BAO and Supernova observations \textbf{(b)}. The contours correspond to cases where $\chi^\prime$ lifetime is $10^5$s, $10^6$s, or $10^7$s. The bigger contour corresponds to $99 \%$ of CL, while the smaller is related to $68 \%$ of CL.}
    \label{fig:lambdaCDM_case}
\end{figure}

Planck collaboration has reported that $N_{eff}$ and $H_0$ are positively correlated \cite{Planck:2018vyg}. This correlation was explored in \cite{ref:Anchordoqui2021} via likelihood functions. Theoretically speaking, the connection between our mechanism and $H_0$ occurs through \eq{eq:deltaN}. For a set of parameters $f\, m_{\chi^\prime}/m_\chi$ and lifetime,$\tau$, we determine $\Delta N_{eff}$. Using the correlation between $\Delta N_{eff}$ and $H_0$ obtained in \cite{ref:Anchordoqui2021}, we exhibit the region of parameter space in terms of $f\, m_{\chi^\prime}/m_\chi$ and $H_0$ for a given lifetime. We do this exercise for two cases. One assuming Planck data only (\fig{fig:lambdaCDM_case_a}), other combining Planck with BAO, and type Ia supernovae data (\fig{fig:lambdaCDM_case_b}). In these two plots the $\Lambda$CDM model was assumed, \Deivid{the cosmological (Planck and BAO) and astrophysical (Ia supernovae) data are taken from \cite{ref:Anchordoqui2021}}. Thus we solidly conclude that we cannot obtain $H_0>71 kms^{-1}Mpc^{-1}$ adopting the $\Lambda$CDM as a prior. One needs to go beyond the $\Lambda$CDM model to find values of $H_0$ consistent with local measurements.

An important observation is that in \fig{fig:lambdaCDM_case} we do not contemplate a non-flat universe, because the curvature does not ameliorate Hubble tension \cite{ref:rezaei2020}.

As expected from \eq{eq:deltaN}, the larger the lifetime the smaller the ratio $f\, m_{\chi^\prime}/m_\chi$ to keep the same $\Delta N_{eff}$. Obviously, this linear relation is a bit lost  with $H_0$, when we factor in the positive correlation between $N_{eff}$ and $H_0$ which is not linear. As we cannot reconcile CMB and local measurements of $H_0$ within the $\Lambda$CDM we will work on a non-standard cosmological background further.

\subsection*{Case 2: Phantom-like Cosmology }

\begin{figure}[htb!]
    \centering
    \subfigure[]{    
        \includegraphics[width=\columnwidth]{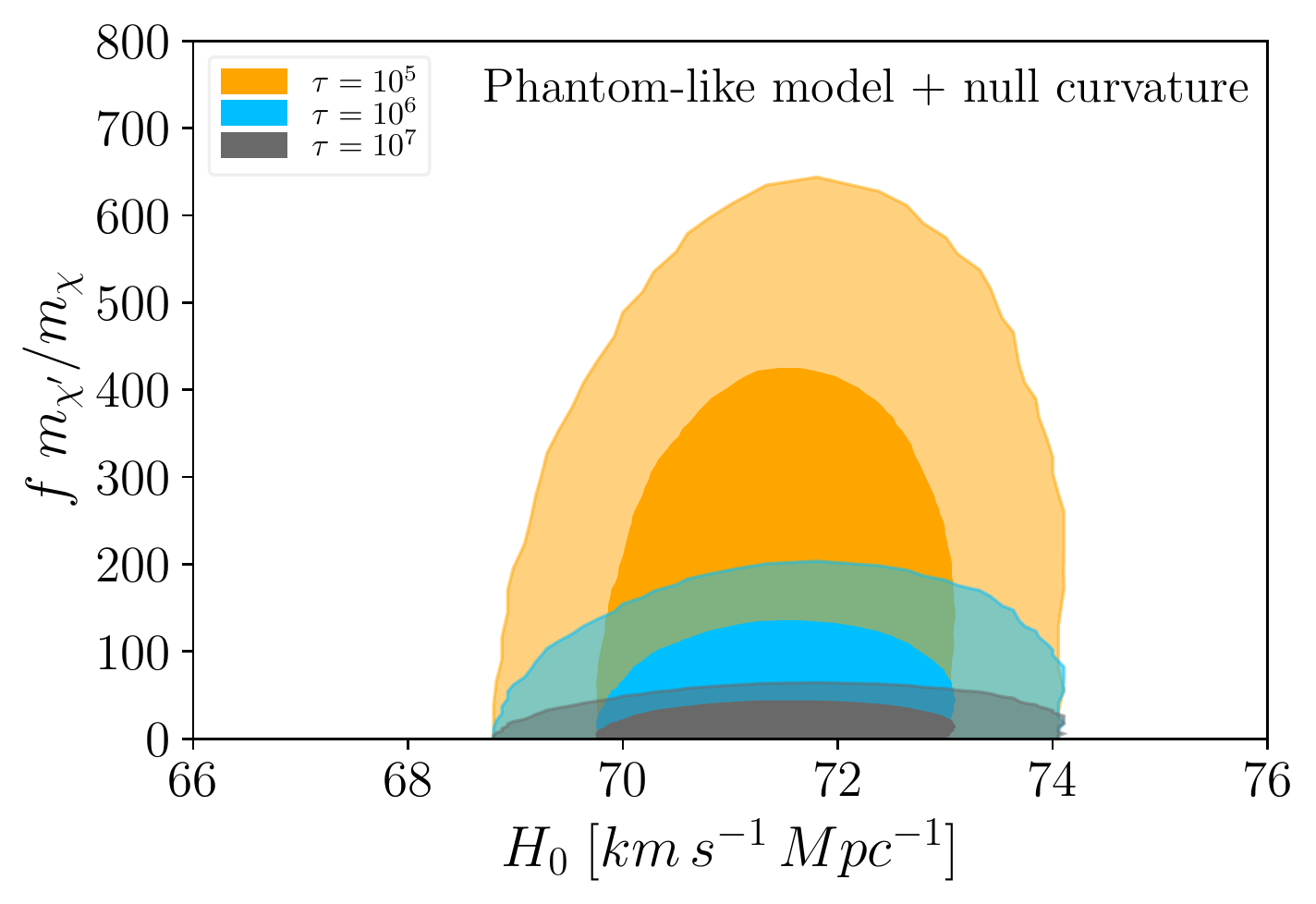}
        \label{fig:NS_case_a}
    }
    \subfigure[]{
        \includegraphics[width=\columnwidth]{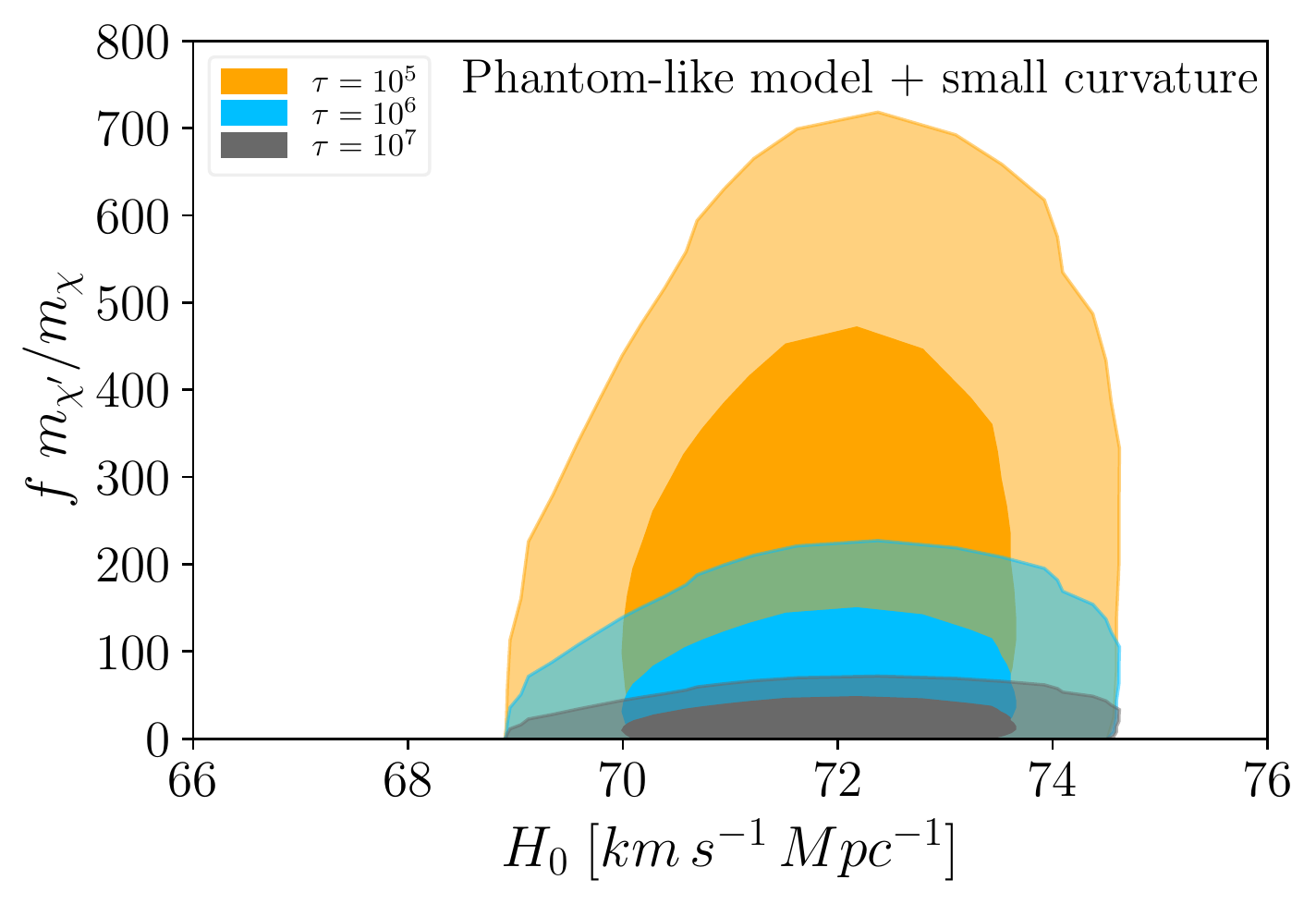}
        \label{fig:NS_case_b}
    }
    \caption{Connection between our model and the value of Hubble constant in phantom-like cases. \textbf{(a)} The contours correspond to cases where $\chi^\prime$ lifetime is $10^5$s, $10^6$s, or $10^7$s. The bigger contour corresponds to $99 \%$ of CL, while the smaller is related to $68 \%$ of CL. It considers a universe with phantom-like quintessence and $\Delta N_{eff}$ in cosmology with null curvature. The bounds were built using Planck 2018 CMB data, BAO, and type Ia data from the Pantheon sample. \textbf{(b)} This case also considers a universe with phantom-like quintessence and $\Delta N_{eff}$, but in this case, a small curvature is added.}
    \label{fig:NS_case}
\end{figure}

We will assume from now on that our cosmological background is a quintessence model. Quintessence is an alternative way to explain the accelerated expansion rate of the universe. It is built on the existence of a scalar field that obeys the equation of state $P = w\rho$, where $P$ are the pressure, $\rho$ is the energy density of quintessence fluid, and $w$ is a real number \cite{ref:weinberg2008cosmology}. The class of models with $w < -1$ are called phantom energy models \cite{Caldwell:1999ew,Caldwell:2003vq,Nojiri:2005sx}. Within this framework, we will assume two scenarios: (i) null curvature $k=0$ \Deivid{and equation of state $P = - 1.004^{+0.038}_{-0.016} \times \rho$}; (ii) non-zero curvature $k>0$ \Deivid{and equation of state $-1.06 \rho < P < -\rho$}. Our reasoning behind these assumptions is the need to change the equation of state of the dark energy fluid to allow larger values for $H_0$ in the fit of the CMB data. The likelihood analyses of these two setups have been carried out and are labeled as $P_7$ and $P_{18}$ in \cite{ref:Anchordoqui2021}. We have checked that these two realizations do not appreciably alter the matter-radiation equality. Thus, \eq{eq:0.16} is still valid as well as our connection between $\Delta N_{eff}$ and $H_0$. 

Similarly, we display the correlation between the parameters of our mechanism and $H_0$ taking $k=0$ (null curvature) in \fig{fig:NS_case_a}, and $k\neq 0$ (non-zero curvature) in \fig{fig:NS_case_b} . \Deivid{We plot them in a similar vein to the previous case: first we use the correlation between $N_{eff}$ and $H_0$ expressed in \cite{ref:Anchordoqui2021}; then we apply this data in \eq{eq:deltaN} for fixed values of $\chi^\prime$ lifetime to relate $fm_{\chi^\prime}/m_{\chi}$ with $H_0$.} It is clear that we easily find $H_0> 71\, {\rm km s^{-1} Mpc^{-1}}$ for $f\, m_\chi^\prime/m_\chi > 100$ and  $\tau \sim 10^5$~s. The difference between null curvature to non-zero curvature is mild. Comparing both plots, we can see that larger values of $f\, m_{\chi^\prime}/m_{\chi}$ are allowed when we go from null curvature to non-zero curvature.  This is expected because with $k\neq 0$ the Hubble rate grows a bit faster. Therefore, the same amount of dark radiation the $k\neq 0$ solution leads to a larger $H_0$.

Now we have shown the region of parameter space in which our mechanism yield a $H_0$ sufficiently large to reconcile CMB and local measurements, we discuss the most important constraints.

\section{BBN constraints}

\begin{figure*}[ht!]
    \centering
    \subfigure[]{
        \includegraphics[width=\columnwidth]{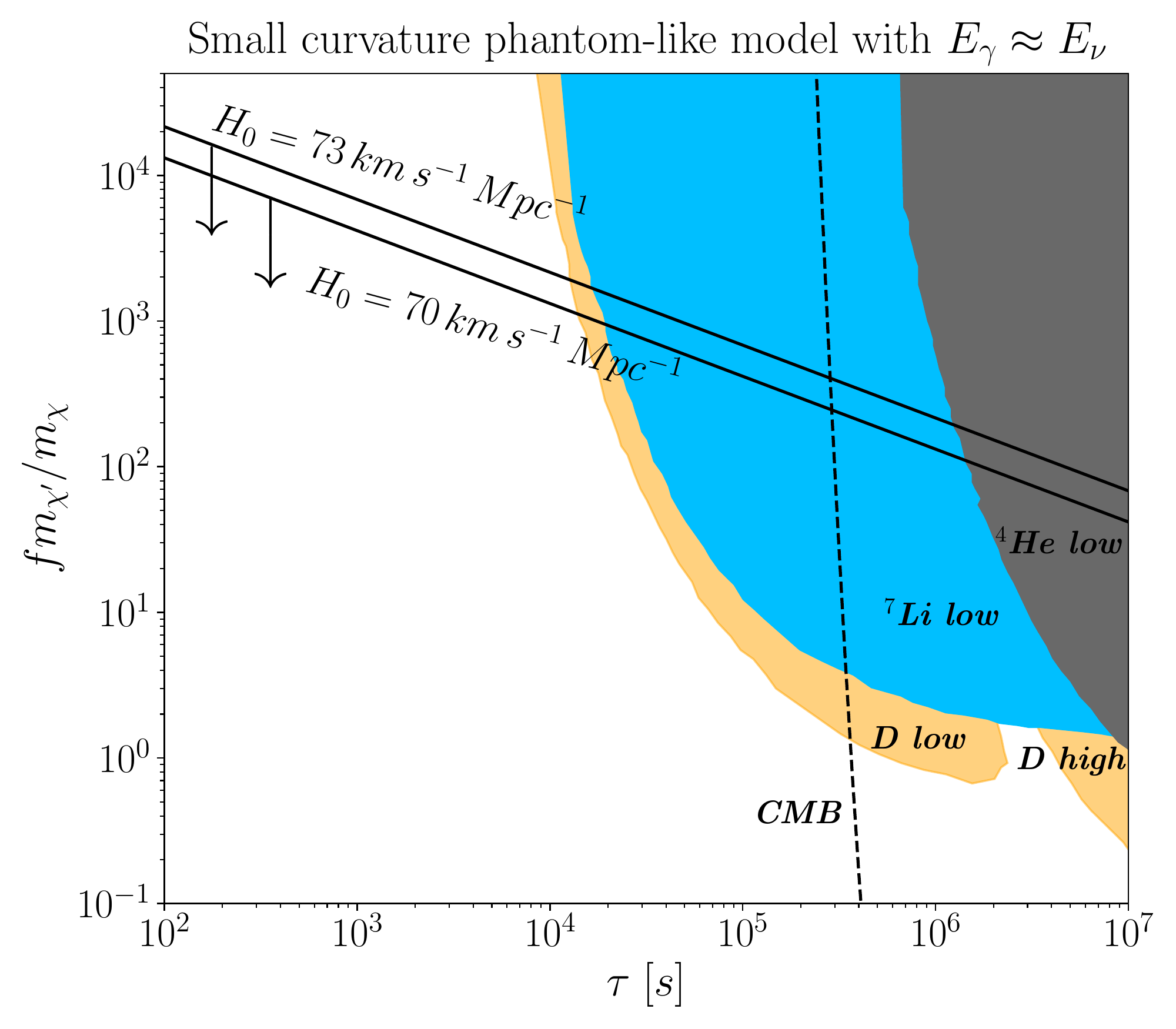}
        \label{fig:BBN_constraints_a}
    }
    \subfigure[]{
        \includegraphics[width=\columnwidth]{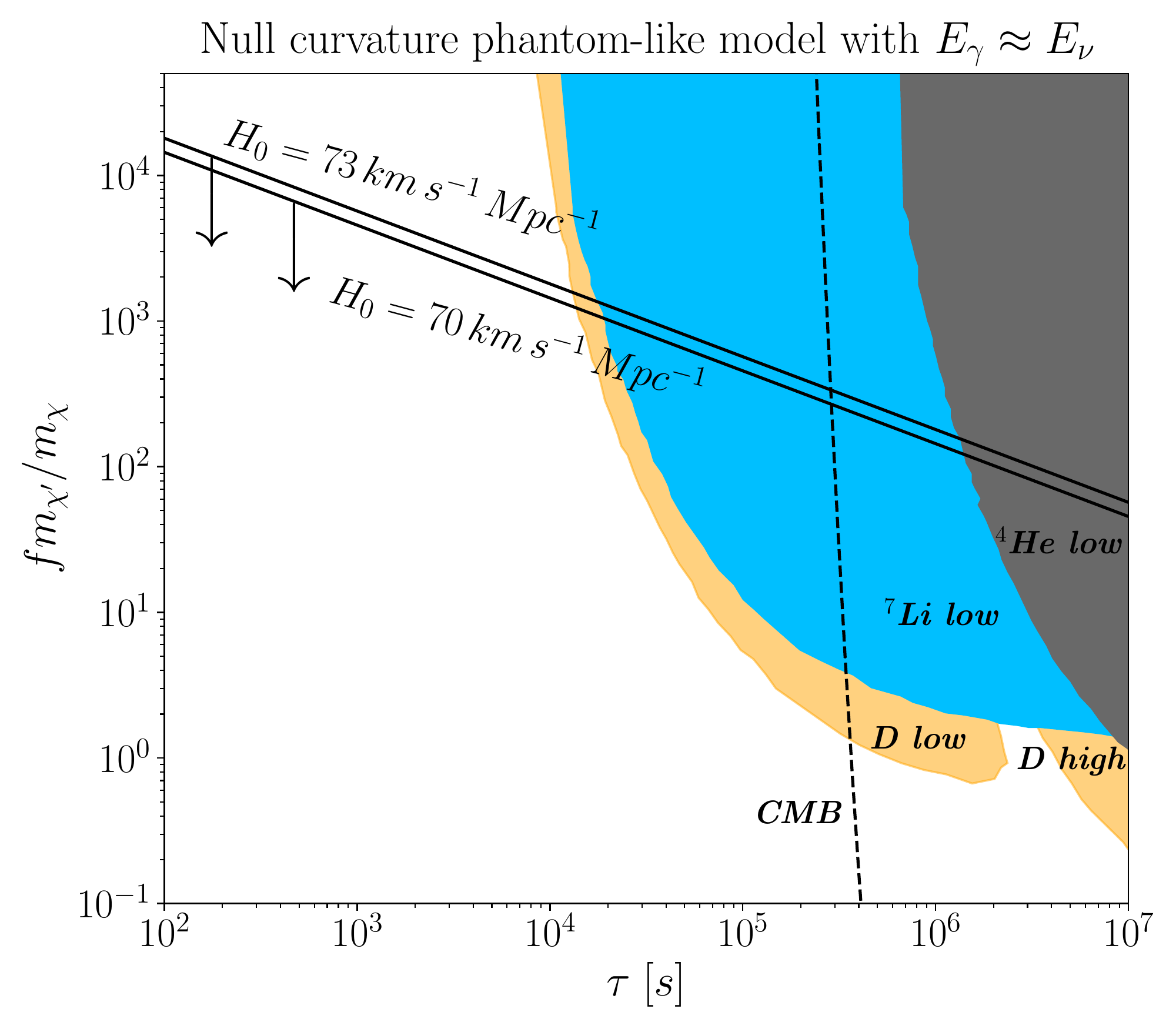}
        \label{fig:BBN_constraints_b}
    }
    \caption{BBN bounds based on light element abundances, and CMB constraint stemming from spectral distortion of the CMB are presented. We overlay the theoretical prediction for $H_0$ using our non-thermal dark matter production mechanism.  In Fig.\ref{fig:BBN_constraints_a} we display the results for $E_\gamma =E_\nu$ and $k\neq 0$, where $E_\gamma$ is the energy of the gamma-rays produced after the $\chi^\prime \rightarrow \chi +\nu$ decay. In Fig.\ref{fig:BBN_constraints_b} we show the results for $E_\gamma =E_\nu$, but with $k= 0$. See text for details.}
    \label{fig:BBN_constraints}
\end{figure*}

The Big Bang Nucleosynthesis is one of the landmarks of early universe cosmology. Any energy injection episode that happens around BBN times may alter the BBN predictions which are consistent with astronomical observations. The decay $\chi^\prime \rightarrow \chi + \nu$ can generate a photon cascade as pointed in \cite{ref:Feng2003,ref:Cyburt2003}. These new photons add electromagnetic energy to the cosmological fluid  which can result in the depletion of Helium, Deuterium, etc. Before showing the final results, we review how these bounds are derived. Before $\chi^\prime$ decay, the universe has a background of photons. Therefore, the energy of photons detected in the CMB is the addition of the energy of this photon background and the energy of new photons generated from our mechanism. For that reason, we write the mean energy of CMB photons as,
\begin{equation}
    E^{CMB}_{\gamma} = E^{BG}_{\gamma}\left( \frac{n^{BG}_{\gamma}}{n^{CMB}_{\gamma}} \right) + E_{\gamma}\left( \frac{n_{\gamma}}{n^{CMB}_{\gamma}} \right),
\end{equation}where $E^{BG}_{\gamma}$ is the mean energy of background photons, $E_{\gamma}$ the mean energy of photons due to the $\chi^\prime$ decay, $n^{BG}_{\gamma}$ the number density of background photons, $n^{CMB}_\gamma$ the number density of CMB photons, and $n_\gamma$ the number density of photons generated by our formalism.

This relation motivates us to define the electromagnetic energy released by $\chi^\prime$ decay as,
\begin{equation}
    \zeta_{EM} \equiv E_{\gamma}Y_{\gamma}.
    \label{eq:zeta_definition}
\end{equation}where $Y_{\gamma} = n_{\gamma}/n^{CMB}_{\gamma}$. 

This equation provides us with a way to calculate the electromagnetic energy introduced by the $\chi^\prime$ decay. Kinematics gives us $E_{\gamma}$, and cosmology the $Y_\gamma$ factor.

Defining the ratio between the dark matter number density and the CMB photons as,
\begin{equation*}
    \begin{split}
        Y_{DM} &\equiv \frac{n_{DM}}{n^{CMB}_{\gamma}} = \frac{n_{CDM} + n_{HDM}}{n^{CMB}_{\gamma}}\\
        &= \frac{n_{CDM}}{n^{CMB}_{\gamma}} \left( 1+f \right).
    \end{split}
\end{equation*}we conclude that is natural to define,
\begin{equation}
    Y_{\chi} \equiv \frac{n_{CDM}}{n^{CMB}_{\gamma}} \times f.
\end{equation}

Using the definition of critical density $(\rho_c \equiv 3H/(8\pi G))$, the definition of density parameter $(\Omega \equiv \rho/\rho_c)$, the cold particle energy density $(\rho= nm)$, and the time evolution of number density of CMB photons $( n^{CMB}_\gamma = n^{CMB}_{\gamma,0}/a^3 )$ \cite{ref:hobson2006GR}, we write $Y_{\chi}$ as,
\begin{equation}
    Y_{\chi} \equiv \frac{n_{CDM}}{n^{CMB}_{\gamma}} \times f = \frac{f}{m_\chi n^{CMB}_{\gamma,0}} \times \Omega_{CDM}a^3 \rho_c.
\end{equation}

As $\Omega_{CDM} = \Omega_{CDM,0}(H_0/H)^2/a^3$, with $\rho_c/\rho_{c,0} = (H/H_0)^2$ \cite{ref:hobson2006GR}, we get,
\begin{equation}
    \Omega_{CDM}a^3 \rho_c = \Omega_{CDM,0} \rho_{c,0},
\end{equation}and consequently,
\begin{equation}
    Y_{\chi} = \frac{f}{m_\chi n^{CMB}_{\gamma,0}} \times \Omega_{CDM,0} \rho_{c,0}.
\end{equation}

With $\rho_{c,0} \approx 1.05 \times 10^{-5}h^2 GeV/cm^3$, $n^{CMB}_{\gamma,0} = 411 cm^{-3}$, and $\Omega_{CDM,0}h^2 = 0.12$ we obtain, 
\begin{equation}
    Y_{\chi} = 3.01 \times 10^{-9} \left( \frac{GeV}{m_{\chi}} \right) \times f.
\end{equation}

The decay $\chi^\prime \rightarrow \chi + \nu$ implies that $n_{\chi^\prime} = n_{\chi} = n_{\nu}$, where $n_\nu$ is not the total neutrinos number density, it is the number density of neutrinos included in the universe due to the $\chi^\prime$ decay. The $\chi^\prime$ decay generates neutrinos that can interact with particles in the background resulting into high-energy photons which induce nuclear reactions and consequently alter the BBN predictions. We will  adopt  $n_{\nu} \approx n_{\gamma}$, which gives in $Y_\gamma \approx Y_\chi$. Conservation of momentum $(p_{\chi^\prime} - p_\nu = p_\chi)$ implies,
\begin{equation}
    E_\nu = \frac{m_{\chi^\prime}}{2}\left[ 1 + \left( \frac{m_\nu}{m_{\chi^\prime}} \right)^2 - \left( \frac{m_\chi}{m_{\chi^\prime}} \right)^2 \right].
\end{equation}
Hence, in the limit where $m_{\chi^\prime} \gg m_{\chi}$, we get $E_\nu = m_{\chi^\prime}/2$. Assuming that all neutrino energy converts into electromagnetic radiation, we obtain $E_\gamma \approx E_\nu$. Thus, 
\begin{equation}
    \zeta_{EM} = E_\nu Y_\chi = 1.5 \times 10^{-9} GeV \times \left( f \frac{m_{\chi^\prime}}{m_\chi} \right).
    \label{eqzeta}
\end{equation}

Knowing how the energetic photons can destroy the light element abundances as derived in the BBN code presented in  \cite{Kawasaki:2017bqm}, we can take this result in terms of energy injection and translate it to our framework as we know from Eq.\eqref{eqzeta} the amount of radiation injected in our non-thermal production mechanism. We overlay these bounds on our findings in \fig{fig:BBN_constraints}. The shaded regions are excluded for either destroying Helium-4, Lithium-7 and Deuterium or inducing a nuclear reaction that saturates the production of Deuterium is dissagreement with astronomical observations \cite{Holtmann:1998gd,Kawasaki:2004yh,Kawasaki:2004qu,Kohri:2006cn,Kawasaki:2007xb,Kawasaki:2008qe,Jittoh:2011ni}.

\section{CMB Bounds}

The injection of electromagnetic energy may also distort the frequency dependence of the CMB spectrum. Double Compton scattering ($\gamma e^-\rightarrow \gamma\gamma e^-$), and bremsstrahlung ($e^-X \rightarrow e^-X\gamma$) are not very efficient at the lifetime we interested in $\tau > 10^4$s. The CMB spectrum as a result relaxes a Bose-Einstein distribution function with chemical potential different from zero. The change in the chemical potential is linked to the lifetime and electromagnetic energy released in the decay process \cite{Feng:2003uy}. Therefore, we plot in the plane $f m_{\chi^\prime}/m_{\chi} \times \tau$ the CMB bound. The limit is delimited by a dashed line in \fig{fig:BBN_constraints}.

\section{Structure Formation}

Now we will justify why the fraction of dark matter particles produces via this non-thermal mechanism should be small using input from structure formation.
The scaling of the free-streaming distance  of a given particle is  understood in terms of the Jeans wavenumber,

\begin{equation}
k_{fs} =\sqrt{\frac{3}{2}} \frac{a H(a)}{v_{med}(a)},
\end{equation}where for $k > k_{fs}$, the density perturbation is damped. The correlation of the galaxy distribution probes the matter power spectrum on scales of  $0.02 h Mpc^{-1} < k < 0.2 h Mpc^{-1}$ at $z \sim 0$ \cite{Zhao:2012xw}. There are other probes such as the Lyman-$\alpha$ spectrum that covers smaller scales \cite{Palanque-Delabrouille:2013gaa}. Using galaxy clustering observations one can assess the maximum amount of hot dark matter in the universe. This hot dark matter component is interpreted in terms of massive neutrinos whose is $\Omega_\nu h^2= \sum m_\nu/94$~eV. The limit is often quoted as $\sum m_\nu \lesssim 0.1$, which implies $\Omega_{HDM} /\Omega_{CDM} \lesssim 0.01$, where we used $\Omega_{CDM}h^2=0.11$.  In other words, $f \lesssim 0.01$ to be consistent with structure formation studies. In more complex dark sector constructions, the presence of non-thermal production mechanism of dark matter is natural.  Notice that even if this non-thermal production be insignificant for overall dark matter energy density, it can give rise to interesting cosmological implications, such as increase $H_0$.

\section{\label{R} Discussions}

Looking at the \fig{fig:BBN_constraints} we conclude that our mechanism can increase the $H_0$ inferred from CMB, and thus reconcile its value with local measurements. We highlight this was only possible assuming phantom-like cosmologies, because within the $\Lambda$CDM model, one cannot solve the $H_0$ problem via $N_{eff}$. As this mechanism represents an energy injection episode, there are restrictive BBN and CMB bounds arise, with BBN being much more severe though. Those constraints left us with a region of parameter where the $\chi^\prime \rightarrow \chi +\nu$ decay process happens between $10^2\, {\rm s} \leq \tau \leq 10^4$~s, for $f m_{\chi^\prime}/m_\chi \sim 10^3 - 10^4 $. Concerning our choice for the $\chi^\prime \rightarrow \chi + \nu$ decay process, it is motivated by model building constructions in the context of supersymmetry and extended gauge sectors, where this decay process is present \cite{Feng:2004zu,Kelso:2013nwa}.\\

We would like to stress that there are alternative explanations for the $H_0$ tension based on different dark energy models. For instance, in \cite{Rezaei:2019xwo}, the authors consider the dark energy density as dynamical, appearing as a power series expansion of the Hubble rate. The idea does not completely solve the $H_0$ problem though, but it alleviates the tension. In \cite{Rezaei:2022bkb}, the authors comprehensively compare different types of dynamical dark energy models that can reduce the $H_0$ discrepancy. Despite the interesting aspects of these papers, our approach is rather orthogonal. We do rely on a dark energy component different from the $\Lambda$CDM. Conversely to the previous references, and others therein, our findings are tied to the dark matter density, and to the production mechanism of dark matter particles, rendering our idea novel in that regard. Hence, we advocate that our solution to the $H_0$ trouble is more appealing because it lies at the interface between particle physics and cosmology, giving rise to a rich phenomenology, and it shows that going beyond the standard thermal production of dark matter leads to a new road into the cosmos, particularly the expansion rate of the universe.  We highlight that in the dark matter literature there is an ongoing discussion about new production mechanisms of dark matter particles. Our work goes precisely in that direction, but with the benefit of solving the $H_0$ tension.

\section{\label{dc} conclusions}

We explored the interplay between particle physics and phantom-like cosmologies to solve the $H_0$ problem via a non-thermal production mechanism of dark matter. If only a fraction of dark matter,$\chi$, is produced via the $\chi^\prime \rightarrow \chi +\nu$ decay process, its non-thermal production can mimic the effect of an extra neutrino species, i.e., a dark radiation. The neutrino species appearing in the final state is a mere choice and does not impact our overal conclusions. If the $\chi^\prime$ particle is sufficiently longed lived, for  $\tau= 10^2-10^4$ sec, this framework can increase $H_0$, but only with the help from phantom-like cosmologies it reaches $H_0\sim 72-74 {\rm kms^{-1}Mpc^{-1}}$ in agreement with local measurements. Our work, shows that the $H_0$ can be troubled by dark matter particles, and it offers us an opportunity to probe the production mechanism of dark matter particles. 

\acknowledgments

FSQ thanks Manfred Lindener for fruitful discussions and the Max Planck Institute fur Kernphysik for the hospitality during the final stages of this work. FSQ is supported by  ICTP-SAIFR FAPESP grants 2016/01343-7 and 2021/14335-0, FAPESP grant 2021/00449-4, CNPq grant 307130/2021-5, Serrapilheira Foundation (grant number Serra - 1912 – 31613). FSQ also acknowledges support from  ANID$-$Millennium Program$-$ICN2019\_044 (Chile).  D. R. da Silva thanks for the support of Coordenaç\~ao de Aperfeiçoamento de Pessoal de N\'ivel Superior (CAPES) under the grant 88887.462084/2019-00. The author JPN acknowledges the support from CAPES under grant 88887.670047/2022-00.

\section*{Data Availability}
The data that connect the Hubble constant \Deivid{$(H_0)$} and the effective number of relativistic particles \Deivid{$(N_{eff})$} analysed during this study are included in reference \cite{ref:Anchordoqui2021}. \Deivid{There, the authors use Planck 2018, BAO and type Ia supernovae data to derive the allowed parameter space in many cosmological cases. In this reference, the $\Lambda$CDM model with $N_{eff}$ is labeled as $\mathcal{P}_1$ case, the phantom-like cosmology with null-curvature is called $\mathcal{P}_7$, and the phantom-like model with small curvature is denoted $\mathcal{P}_{18}$.}

The data that provides bounds from BBN and CMB used during this study are included in reference \cite{Feng:2003uy}.

\bibliography{references.bib}

\begin{thebibliography}{40}%
\makeatletter
\providecommand \@ifxundefined [1]{%
 \@ifx{#1\undefined}
}%
\providecommand \@ifnum [1]{%
 \ifnum #1\expandafter \@firstoftwo
 \else \expandafter \@secondoftwo
 \fi
}%
\providecommand \@ifx [1]{%
 \ifx #1\expandafter \@firstoftwo
 \else \expandafter \@secondoftwo
 \fi
}%
\providecommand \natexlab [1]{#1}%
\providecommand \enquote  [1]{``#1''}%
\providecommand \bibnamefont  [1]{#1}%
\providecommand \bibfnamefont [1]{#1}%
\providecommand \citenamefont [1]{#1}%
\providecommand \href@noop [0]{\@secondoftwo}%
\providecommand \href [0]{\begingroup \@sanitize@url \@href}%
\providecommand \@href[1]{\@@startlink{#1}\@@href}%
\providecommand \@@href[1]{\endgroup#1\@@endlink}%
\providecommand \@sanitize@url [0]{\catcode `\\12\catcode `\$12\catcode
  `\&12\catcode `\#12\catcode `\^12\catcode `\_12\catcode `\%12\relax}%
\providecommand \@@startlink[1]{}%
\providecommand \@@endlink[0]{}%
\providecommand \url  [0]{\begingroup\@sanitize@url \@url }%
\providecommand \@url [1]{\endgroup\@href {#1}{\urlprefix }}%
\providecommand \urlprefix  [0]{URL }%
\providecommand \Eprint [0]{\href }%
\providecommand \doibase [0]{http://dx.doi.org/}%
\providecommand \selectlanguage [0]{\@gobble}%
\providecommand \bibinfo  [0]{\@secondoftwo}%
\providecommand \bibfield  [0]{\@secondoftwo}%
\providecommand \translation [1]{[#1]}%
\providecommand \BibitemOpen [0]{}%
\providecommand \bibitemStop [0]{}%
\providecommand \bibitemNoStop [0]{.\EOS\space}%
\providecommand \EOS [0]{\spacefactor3000\relax}%
\providecommand \BibitemShut  [1]{\csname bibitem#1\endcsname}%
\let\auto@bib@innerbib\@empty
\bibitem [{\citenamefont {Aghanim}\ \emph
  {et~al.}(2020{\natexlab{a}})\citenamefont {Aghanim} \emph
  {et~al.}}]{Aghanim:2018eyx}%
  \BibitemOpen
  \bibfield  {author} {\bibinfo {author} {\bibfnamefont {N.}~\bibnamefont
  {Aghanim}} \emph {et~al.} (\bibinfo {collaboration} {Planck}),\ }\href
  {\doibase 10.1051/0004-6361/201833910} {\bibfield  {journal} {\bibinfo
  {journal} {Astron. Astrophys.}\ }\textbf {\bibinfo {volume} {641}},\ \bibinfo
  {pages} {A6} (\bibinfo {year} {2020}{\natexlab{a}})},\ \Eprint
  {http://arxiv.org/abs/1807.06209} {arXiv:1807.06209 [astro-ph.CO]}
  \BibitemShut {NoStop}%
\bibitem [{\citenamefont {Ade}\ \emph {et~al.}(2016)\citenamefont {Ade} \emph
  {et~al.}}]{Planck:2015fie}%
  \BibitemOpen
  \bibfield  {author} {\bibinfo {author} {\bibfnamefont {P.~A.~R.}\
  \bibnamefont {Ade}} \emph {et~al.} (\bibinfo {collaboration} {Planck}),\
  }\href {\doibase 10.1051/0004-6361/201525830} {\bibfield  {journal} {\bibinfo
   {journal} {Astron. Astrophys.}\ }\textbf {\bibinfo {volume} {594}},\
  \bibinfo {pages} {A13} (\bibinfo {year} {2016})},\ \Eprint
  {http://arxiv.org/abs/1502.01589} {arXiv:1502.01589 [astro-ph.CO]}
  \BibitemShut {NoStop}%
\bibitem [{\citenamefont {Aghanim}\ \emph
  {et~al.}(2020{\natexlab{b}})\citenamefont {Aghanim} \emph
  {et~al.}}]{Planck:2018vyg}%
  \BibitemOpen
  \bibfield  {author} {\bibinfo {author} {\bibfnamefont {N.}~\bibnamefont
  {Aghanim}} \emph {et~al.} (\bibinfo {collaboration} {Planck}),\ }\href
  {\doibase 10.1051/0004-6361/201833910} {\bibfield  {journal} {\bibinfo
  {journal} {Astron. Astrophys.}\ }\textbf {\bibinfo {volume} {641}},\ \bibinfo
  {pages} {A6} (\bibinfo {year} {2020}{\natexlab{b}})},\ \bibinfo {note}
  {[Erratum: Astron.Astrophys. 652, C4 (2021)]},\ \Eprint
  {http://arxiv.org/abs/1807.06209} {arXiv:1807.06209 [astro-ph.CO]}
  \BibitemShut {NoStop}%
\bibitem [{\citenamefont {Aiola}\ \emph {et~al.}(2020)\citenamefont {Aiola}
  \emph {et~al.}}]{ACT:2020gnv}%
  \BibitemOpen
  \bibfield  {author} {\bibinfo {author} {\bibfnamefont {S.}~\bibnamefont
  {Aiola}} \emph {et~al.} (\bibinfo {collaboration} {ACT}),\ }\href {\doibase
  10.1088/1475-7516/2020/12/047} {\bibfield  {journal} {\bibinfo  {journal}
  {JCAP}\ }\textbf {\bibinfo {volume} {12}},\ \bibinfo {pages} {047} (\bibinfo
  {year} {2020})},\ \Eprint {http://arxiv.org/abs/2007.07288} {arXiv:2007.07288
  [astro-ph.CO]} \BibitemShut {NoStop}%
\bibitem [{\citenamefont {Balkenhol}\ \emph {et~al.}(2021)\citenamefont
  {Balkenhol} \emph {et~al.}}]{SPT-3G:2021wgf}%
  \BibitemOpen
  \bibfield  {author} {\bibinfo {author} {\bibfnamefont {L.}~\bibnamefont
  {Balkenhol}} \emph {et~al.} (\bibinfo {collaboration} {SPT-3G}),\ }\href
  {\doibase 10.1103/PhysRevD.104.083509} {\bibfield  {journal} {\bibinfo
  {journal} {Phys. Rev. D}\ }\textbf {\bibinfo {volume} {104}},\ \bibinfo
  {pages} {083509} (\bibinfo {year} {2021})},\ \Eprint
  {http://arxiv.org/abs/2103.13618} {arXiv:2103.13618 [astro-ph.CO]}
  \BibitemShut {NoStop}%
\bibitem [{\citenamefont {Valentino}\ \emph {et~al.}(2021)\citenamefont
  {Valentino}, \citenamefont {Mena}, \citenamefont {Pan}, \citenamefont
  {Visinelli}, \citenamefont {Yang}, \citenamefont {Melchiorri}, \citenamefont
  {Mota}, \citenamefont {Riess},\ and\ \citenamefont {Silk}}]{DiValentino2021}%
  \BibitemOpen
  \bibfield  {author} {\bibinfo {author} {\bibfnamefont {E.~D.}\ \bibnamefont
  {Valentino}}, \bibinfo {author} {\bibfnamefont {O.}~\bibnamefont {Mena}},
  \bibinfo {author} {\bibfnamefont {S.}~\bibnamefont {Pan}}, \bibinfo {author}
  {\bibfnamefont {L.}~\bibnamefont {Visinelli}}, \bibinfo {author}
  {\bibfnamefont {W.}~\bibnamefont {Yang}}, \bibinfo {author} {\bibfnamefont
  {A.}~\bibnamefont {Melchiorri}}, \bibinfo {author} {\bibfnamefont {D.~F.}\
  \bibnamefont {Mota}}, \bibinfo {author} {\bibfnamefont {A.~G.}\ \bibnamefont
  {Riess}}, \ and\ \bibinfo {author} {\bibfnamefont {J.}~\bibnamefont {Silk}},\
  }\href {\doibase 10.1088/1361-6382/ac086d} {\bibfield  {journal} {\bibinfo
  {journal} {Classical and Quantum Gravity}\ }\textbf {\bibinfo {volume}
  {38}},\ \bibinfo {pages} {153001} (\bibinfo {year} {2021})}\BibitemShut
  {NoStop}%
\bibitem [{\citenamefont {Kenworthy}\ \emph {et~al.}(2022)\citenamefont
  {Kenworthy}, \citenamefont {Riess}, \citenamefont {Scolnic}, \citenamefont
  {Yuan}, \citenamefont {Bernal}, \citenamefont {Brout}, \citenamefont
  {Cassertano}, \citenamefont {Jones}, \citenamefont {Macri},\ and\
  \citenamefont {Peterson}}]{Kenworthy:2022jdh}%
  \BibitemOpen
  \bibfield  {author} {\bibinfo {author} {\bibfnamefont {W.~D.}\ \bibnamefont
  {Kenworthy}}, \bibinfo {author} {\bibfnamefont {A.~G.}\ \bibnamefont
  {Riess}}, \bibinfo {author} {\bibfnamefont {D.}~\bibnamefont {Scolnic}},
  \bibinfo {author} {\bibfnamefont {W.}~\bibnamefont {Yuan}}, \bibinfo {author}
  {\bibfnamefont {J.~L.}\ \bibnamefont {Bernal}}, \bibinfo {author}
  {\bibfnamefont {D.}~\bibnamefont {Brout}}, \bibinfo {author} {\bibfnamefont
  {S.}~\bibnamefont {Cassertano}}, \bibinfo {author} {\bibfnamefont {D.~O.}\
  \bibnamefont {Jones}}, \bibinfo {author} {\bibfnamefont {L.}~\bibnamefont
  {Macri}}, \ and\ \bibinfo {author} {\bibfnamefont {E.}~\bibnamefont
  {Peterson}},\ }\href@noop {} {\  (\bibinfo {year} {2022})},\ \Eprint
  {http://arxiv.org/abs/2204.10866} {arXiv:2204.10866 [astro-ph.CO]}
  \BibitemShut {NoStop}%
\bibitem [{\citenamefont {Anchordoqui}\ \emph {et~al.}(2021)\citenamefont
  {Anchordoqui}, \citenamefont {Di~Valentino}, \citenamefont {Pan},\ and\
  \citenamefont {Yang}}]{ref:Anchordoqui2021}%
  \BibitemOpen
  \bibfield  {author} {\bibinfo {author} {\bibfnamefont {L.~A.}\ \bibnamefont
  {Anchordoqui}}, \bibinfo {author} {\bibfnamefont {E.}~\bibnamefont
  {Di~Valentino}}, \bibinfo {author} {\bibfnamefont {S.}~\bibnamefont {Pan}}, \
  and\ \bibinfo {author} {\bibfnamefont {W.}~\bibnamefont {Yang}},\ }\href
  {\doibase 10.1016/j.jheap.2021.08.001} {\bibfield  {journal} {\bibinfo
  {journal} {JHEAp}\ }\textbf {\bibinfo {volume} {32}},\ \bibinfo {pages} {28}
  (\bibinfo {year} {2021})},\ \Eprint {http://arxiv.org/abs/2107.13932}
  {arXiv:2107.13932 [astro-ph.CO]} \BibitemShut {NoStop}%
\bibitem [{\citenamefont {Shah}\ \emph {et~al.}(2021)\citenamefont {Shah},
  \citenamefont {Lemos},\ and\ \citenamefont {Lahav}}]{Shah:2021onj}%
  \BibitemOpen
  \bibfield  {author} {\bibinfo {author} {\bibfnamefont {P.}~\bibnamefont
  {Shah}}, \bibinfo {author} {\bibfnamefont {P.}~\bibnamefont {Lemos}}, \ and\
  \bibinfo {author} {\bibfnamefont {O.}~\bibnamefont {Lahav}},\ }\href
  {\doibase 10.1007/s00159-021-00137-4} {\bibfield  {journal} {\bibinfo
  {journal} {Astron. Astrophys. Rev.}\ }\textbf {\bibinfo {volume} {29}},\
  \bibinfo {pages} {9} (\bibinfo {year} {2021})},\ \Eprint
  {http://arxiv.org/abs/2109.01161} {arXiv:2109.01161 [astro-ph.CO]}
  \BibitemShut {NoStop}%
\bibitem [{\citenamefont {Di~Valentino}\ \emph
  {et~al.}(2021{\natexlab{a}})\citenamefont {Di~Valentino}, \citenamefont
  {Mena}, \citenamefont {Pan}, \citenamefont {Visinelli}, \citenamefont {Yang},
  \citenamefont {Melchiorri}, \citenamefont {Mota}, \citenamefont {Riess},\
  and\ \citenamefont {Silk}}]{DiValentino:2021izs}%
  \BibitemOpen
  \bibfield  {author} {\bibinfo {author} {\bibfnamefont {E.}~\bibnamefont
  {Di~Valentino}}, \bibinfo {author} {\bibfnamefont {O.}~\bibnamefont {Mena}},
  \bibinfo {author} {\bibfnamefont {S.}~\bibnamefont {Pan}}, \bibinfo {author}
  {\bibfnamefont {L.}~\bibnamefont {Visinelli}}, \bibinfo {author}
  {\bibfnamefont {W.}~\bibnamefont {Yang}}, \bibinfo {author} {\bibfnamefont
  {A.}~\bibnamefont {Melchiorri}}, \bibinfo {author} {\bibfnamefont {D.~F.}\
  \bibnamefont {Mota}}, \bibinfo {author} {\bibfnamefont {A.~G.}\ \bibnamefont
  {Riess}}, \ and\ \bibinfo {author} {\bibfnamefont {J.}~\bibnamefont {Silk}},\
  }\href {\doibase 10.1088/1361-6382/ac086d} {\bibfield  {journal} {\bibinfo
  {journal} {Class. Quant. Grav.}\ }\textbf {\bibinfo {volume} {38}},\ \bibinfo
  {pages} {153001} (\bibinfo {year} {2021}{\natexlab{a}})},\ \Eprint
  {http://arxiv.org/abs/2103.01183} {arXiv:2103.01183 [astro-ph.CO]}
  \BibitemShut {NoStop}%
\bibitem [{\citenamefont {Di~Valentino}\ \emph
  {et~al.}(2021{\natexlab{b}})\citenamefont {Di~Valentino} \emph
  {et~al.}}]{DiValentino:2020zio}%
  \BibitemOpen
  \bibfield  {author} {\bibinfo {author} {\bibfnamefont {E.}~\bibnamefont
  {Di~Valentino}} \emph {et~al.},\ }\href {\doibase
  10.1016/j.astropartphys.2021.102605} {\bibfield  {journal} {\bibinfo
  {journal} {Astropart. Phys.}\ }\textbf {\bibinfo {volume} {131}},\ \bibinfo
  {pages} {102605} (\bibinfo {year} {2021}{\natexlab{b}})},\ \Eprint
  {http://arxiv.org/abs/2008.11284} {arXiv:2008.11284 [astro-ph.CO]}
  \BibitemShut {NoStop}%
\bibitem [{\citenamefont {Abdalla}\ \emph {et~al.}(2022)\citenamefont {Abdalla}
  \emph {et~al.}}]{Abdalla:2022yfr}%
  \BibitemOpen
  \bibfield  {author} {\bibinfo {author} {\bibfnamefont {E.}~\bibnamefont
  {Abdalla}} \emph {et~al.},\ }\href {\doibase 10.1016/j.jheap.2022.04.002}
  {\bibfield  {journal} {\bibinfo  {journal} {JHEAp}\ }\textbf {\bibinfo
  {volume} {34}},\ \bibinfo {pages} {49} (\bibinfo {year} {2022})},\ \Eprint
  {http://arxiv.org/abs/2203.06142} {arXiv:2203.06142 [astro-ph.CO]}
  \BibitemShut {NoStop}%
\bibitem [{\citenamefont {Hooper}\ \emph {et~al.}(2012)\citenamefont {Hooper},
  \citenamefont {Queiroz},\ and\ \citenamefont {Gnedin}}]{Hooper:2011aj}%
  \BibitemOpen
  \bibfield  {author} {\bibinfo {author} {\bibfnamefont {D.}~\bibnamefont
  {Hooper}}, \bibinfo {author} {\bibfnamefont {F.~S.}\ \bibnamefont {Queiroz}},
  \ and\ \bibinfo {author} {\bibfnamefont {N.~Y.}\ \bibnamefont {Gnedin}},\
  }\href {\doibase 10.1103/PhysRevD.85.063513} {\bibfield  {journal} {\bibinfo
  {journal} {Phys. Rev. D}\ }\textbf {\bibinfo {volume} {85}},\ \bibinfo
  {pages} {063513} (\bibinfo {year} {2012})},\ \Eprint
  {http://arxiv.org/abs/1111.6599} {arXiv:1111.6599 [astro-ph.CO]} \BibitemShut
  {NoStop}%
\bibitem [{\citenamefont {Kelso}\ \emph {et~al.}(2013)\citenamefont {Kelso},
  \citenamefont {Profumo},\ and\ \citenamefont {Queiroz}}]{Kelso:2013paa}%
  \BibitemOpen
  \bibfield  {author} {\bibinfo {author} {\bibfnamefont {C.}~\bibnamefont
  {Kelso}}, \bibinfo {author} {\bibfnamefont {S.}~\bibnamefont {Profumo}}, \
  and\ \bibinfo {author} {\bibfnamefont {F.~S.}\ \bibnamefont {Queiroz}},\
  }\href {\doibase 10.1103/PhysRevD.88.023511} {\bibfield  {journal} {\bibinfo
  {journal} {Phys. Rev. D}\ }\textbf {\bibinfo {volume} {88}},\ \bibinfo
  {pages} {023511} (\bibinfo {year} {2013})},\ \Eprint
  {http://arxiv.org/abs/1304.5243} {arXiv:1304.5243 [hep-ph]} \BibitemShut
  {NoStop}%
\bibitem [{\citenamefont {Allahverdi}\ \emph {et~al.}(2015)\citenamefont
  {Allahverdi}, \citenamefont {Dutta}, \citenamefont {Queiroz}, \citenamefont
  {Strigari},\ and\ \citenamefont {Wang}}]{Allahverdi:2014bva}%
  \BibitemOpen
  \bibfield  {author} {\bibinfo {author} {\bibfnamefont {R.}~\bibnamefont
  {Allahverdi}}, \bibinfo {author} {\bibfnamefont {B.}~\bibnamefont {Dutta}},
  \bibinfo {author} {\bibfnamefont {F.~S.}\ \bibnamefont {Queiroz}}, \bibinfo
  {author} {\bibfnamefont {L.~E.}\ \bibnamefont {Strigari}}, \ and\ \bibinfo
  {author} {\bibfnamefont {M.-Y.}\ \bibnamefont {Wang}},\ }\href {\doibase
  10.1103/PhysRevD.91.055033} {\bibfield  {journal} {\bibinfo  {journal} {Phys.
  Rev. D}\ }\textbf {\bibinfo {volume} {91}},\ \bibinfo {pages} {055033}
  (\bibinfo {year} {2015})},\ \Eprint {http://arxiv.org/abs/1412.4391}
  {arXiv:1412.4391 [hep-ph]} \BibitemShut {NoStop}%
\bibitem [{\citenamefont {Bringmann}\ \emph {et~al.}(2018)\citenamefont
  {Bringmann}, \citenamefont {Kahlhoefer}, \citenamefont {Schmidt-Hoberg},\
  and\ \citenamefont {Walia}}]{Bringmann:2018jpr}%
  \BibitemOpen
  \bibfield  {author} {\bibinfo {author} {\bibfnamefont {T.}~\bibnamefont
  {Bringmann}}, \bibinfo {author} {\bibfnamefont {F.}~\bibnamefont
  {Kahlhoefer}}, \bibinfo {author} {\bibfnamefont {K.}~\bibnamefont
  {Schmidt-Hoberg}}, \ and\ \bibinfo {author} {\bibfnamefont {P.}~\bibnamefont
  {Walia}},\ }\href {\doibase 10.1103/PhysRevD.98.023543} {\bibfield  {journal}
  {\bibinfo  {journal} {Phys. Rev. D}\ }\textbf {\bibinfo {volume} {98}},\
  \bibinfo {pages} {023543} (\bibinfo {year} {2018})},\ \Eprint
  {http://arxiv.org/abs/1803.03644} {arXiv:1803.03644 [astro-ph.CO]}
  \BibitemShut {NoStop}%
\bibitem [{\citenamefont {Aghanim}\ \emph
  {et~al.}(2020{\natexlab{c}})\citenamefont {Aghanim} \emph
  {et~al.}}]{ref:Planck2018}%
  \BibitemOpen
  \bibfield  {author} {\bibinfo {author} {\bibfnamefont {N.}~\bibnamefont
  {Aghanim}} \emph {et~al.} (\bibinfo {collaboration} {Planck}),\ }\href
  {\doibase 10.1051/0004-6361/201833910} {\bibfield  {journal} {\bibinfo
  {journal} {Astron. Astrophys.}\ }\textbf {\bibinfo {volume} {641}},\ \bibinfo
  {pages} {A6} (\bibinfo {year} {2020}{\natexlab{c}})},\ \bibinfo {note}
  {[Erratum: Astron.Astrophys. 652, C4 (2021)]},\ \Eprint
  {http://arxiv.org/abs/1807.06209} {arXiv:1807.06209 [astro-ph.CO]}
  \BibitemShut {NoStop}%
\bibitem [{\citenamefont {Rezaei}\ \emph {et~al.}(2020)\citenamefont {Rezaei},
  \citenamefont {Naderi}, \citenamefont {Malekjani},\ and\ \citenamefont
  {Mehrabi}}]{ref:rezaei2020}%
  \BibitemOpen
  \bibfield  {author} {\bibinfo {author} {\bibfnamefont {M.}~\bibnamefont
  {Rezaei}}, \bibinfo {author} {\bibfnamefont {T.}~\bibnamefont {Naderi}},
  \bibinfo {author} {\bibfnamefont {M.}~\bibnamefont {Malekjani}}, \ and\
  \bibinfo {author} {\bibfnamefont {A.}~\bibnamefont {Mehrabi}},\ }\href@noop
  {} {\bibfield  {journal} {\bibinfo  {journal} {The European Physical Journal
  C}\ }\textbf {\bibinfo {volume} {80}},\ \bibinfo {pages} {1} (\bibinfo {year}
  {2020})}\BibitemShut {NoStop}%
\bibitem [{\citenamefont {Weinberg}(2008)}]{ref:weinberg2008cosmology}%
  \BibitemOpen
  \bibfield  {author} {\bibinfo {author} {\bibfnamefont {S.}~\bibnamefont
  {Weinberg}},\ }\href@noop {} {\emph {\bibinfo {title} {Cosmology}}}\
  (\bibinfo  {publisher} {OUP Oxford},\ \bibinfo {year} {2008})\BibitemShut
  {NoStop}%
\bibitem [{\citenamefont {Caldwell}(2002)}]{Caldwell:1999ew}%
  \BibitemOpen
  \bibfield  {author} {\bibinfo {author} {\bibfnamefont {R.~R.}\ \bibnamefont
  {Caldwell}},\ }\href {\doibase 10.1016/S0370-2693(02)02589-3} {\bibfield
  {journal} {\bibinfo  {journal} {Phys. Lett. B}\ }\textbf {\bibinfo {volume}
  {545}},\ \bibinfo {pages} {23} (\bibinfo {year} {2002})},\ \Eprint
  {http://arxiv.org/abs/astro-ph/9908168} {arXiv:astro-ph/9908168} \BibitemShut
  {NoStop}%
\bibitem [{\citenamefont {Caldwell}\ \emph {et~al.}(2003)\citenamefont
  {Caldwell}, \citenamefont {Kamionkowski},\ and\ \citenamefont
  {Weinberg}}]{Caldwell:2003vq}%
  \BibitemOpen
  \bibfield  {author} {\bibinfo {author} {\bibfnamefont {R.~R.}\ \bibnamefont
  {Caldwell}}, \bibinfo {author} {\bibfnamefont {M.}~\bibnamefont
  {Kamionkowski}}, \ and\ \bibinfo {author} {\bibfnamefont {N.~N.}\
  \bibnamefont {Weinberg}},\ }\href {\doibase 10.1103/PhysRevLett.91.071301}
  {\bibfield  {journal} {\bibinfo  {journal} {Phys. Rev. Lett.}\ }\textbf
  {\bibinfo {volume} {91}},\ \bibinfo {pages} {071301} (\bibinfo {year}
  {2003})},\ \Eprint {http://arxiv.org/abs/astro-ph/0302506}
  {arXiv:astro-ph/0302506} \BibitemShut {NoStop}%
\bibitem [{\citenamefont {Nojiri}\ \emph {et~al.}(2005)\citenamefont {Nojiri},
  \citenamefont {Odintsov},\ and\ \citenamefont {Tsujikawa}}]{Nojiri:2005sx}%
  \BibitemOpen
  \bibfield  {author} {\bibinfo {author} {\bibfnamefont {S.}~\bibnamefont
  {Nojiri}}, \bibinfo {author} {\bibfnamefont {S.~D.}\ \bibnamefont
  {Odintsov}}, \ and\ \bibinfo {author} {\bibfnamefont {S.}~\bibnamefont
  {Tsujikawa}},\ }\href {\doibase 10.1103/PhysRevD.71.063004} {\bibfield
  {journal} {\bibinfo  {journal} {Phys. Rev. D}\ }\textbf {\bibinfo {volume}
  {71}},\ \bibinfo {pages} {063004} (\bibinfo {year} {2005})},\ \Eprint
  {http://arxiv.org/abs/hep-th/0501025} {arXiv:hep-th/0501025} \BibitemShut
  {NoStop}%
\bibitem [{\citenamefont {Feng}\ \emph
  {et~al.}(2003{\natexlab{a}})\citenamefont {Feng}, \citenamefont {Rajaraman},\
  and\ \citenamefont {Takayama}}]{ref:Feng2003}%
  \BibitemOpen
  \bibfield  {author} {\bibinfo {author} {\bibfnamefont {J.~L.}\ \bibnamefont
  {Feng}}, \bibinfo {author} {\bibfnamefont {A.}~\bibnamefont {Rajaraman}}, \
  and\ \bibinfo {author} {\bibfnamefont {F.}~\bibnamefont {Takayama}},\ }\href
  {\doibase 10.1103/PhysRevD.68.063504} {\bibfield  {journal} {\bibinfo
  {journal} {Phys. Rev. D}\ }\textbf {\bibinfo {volume} {68}},\ \bibinfo
  {pages} {063504} (\bibinfo {year} {2003}{\natexlab{a}})},\ \Eprint
  {http://arxiv.org/abs/hep-ph/0306024} {arXiv:hep-ph/0306024} \BibitemShut
  {NoStop}%
\bibitem [{\citenamefont {Cyburt}\ \emph {et~al.}(2003)\citenamefont {Cyburt},
  \citenamefont {Ellis}, \citenamefont {Fields},\ and\ \citenamefont
  {Olive}}]{ref:Cyburt2003}%
  \BibitemOpen
  \bibfield  {author} {\bibinfo {author} {\bibfnamefont {R.~H.}\ \bibnamefont
  {Cyburt}}, \bibinfo {author} {\bibfnamefont {J.~R.}\ \bibnamefont {Ellis}},
  \bibinfo {author} {\bibfnamefont {B.~D.}\ \bibnamefont {Fields}}, \ and\
  \bibinfo {author} {\bibfnamefont {K.~A.}\ \bibnamefont {Olive}},\ }\href
  {\doibase 10.1103/PhysRevD.67.103521} {\bibfield  {journal} {\bibinfo
  {journal} {Phys. Rev. D}\ }\textbf {\bibinfo {volume} {67}},\ \bibinfo
  {pages} {103521} (\bibinfo {year} {2003})},\ \Eprint
  {http://arxiv.org/abs/astro-ph/0211258} {arXiv:astro-ph/0211258} \BibitemShut
  {NoStop}%
\bibitem [{\citenamefont {Hobson}\ \emph {et~al.}(2006)\citenamefont {Hobson},
  \citenamefont {Efstathiou},\ and\ \citenamefont
  {Lasenby}}]{ref:hobson2006GR}%
  \BibitemOpen
  \bibfield  {author} {\bibinfo {author} {\bibfnamefont {M.~P.}\ \bibnamefont
  {Hobson}}, \bibinfo {author} {\bibfnamefont {G.~P.}\ \bibnamefont
  {Efstathiou}}, \ and\ \bibinfo {author} {\bibfnamefont {A.~N.}\ \bibnamefont
  {Lasenby}},\ }\href@noop {} {\emph {\bibinfo {title} {General relativity: an
  introduction for physicists}}}\ (\bibinfo  {publisher} {Cambridge University
  Press},\ \bibinfo {year} {2006})\BibitemShut {NoStop}%
\bibitem [{\citenamefont {Kawasaki}\ \emph {et~al.}(2018)\citenamefont
  {Kawasaki}, \citenamefont {Kohri}, \citenamefont {Moroi},\ and\ \citenamefont
  {Takaesu}}]{Kawasaki:2017bqm}%
  \BibitemOpen
  \bibfield  {author} {\bibinfo {author} {\bibfnamefont {M.}~\bibnamefont
  {Kawasaki}}, \bibinfo {author} {\bibfnamefont {K.}~\bibnamefont {Kohri}},
  \bibinfo {author} {\bibfnamefont {T.}~\bibnamefont {Moroi}}, \ and\ \bibinfo
  {author} {\bibfnamefont {Y.}~\bibnamefont {Takaesu}},\ }\href {\doibase
  10.1103/PhysRevD.97.023502} {\bibfield  {journal} {\bibinfo  {journal} {Phys.
  Rev. D}\ }\textbf {\bibinfo {volume} {97}},\ \bibinfo {pages} {023502}
  (\bibinfo {year} {2018})},\ \Eprint {http://arxiv.org/abs/1709.01211}
  {arXiv:1709.01211 [hep-ph]} \BibitemShut {NoStop}%
\bibitem [{\citenamefont {Holtmann}\ \emph {et~al.}(1999)\citenamefont
  {Holtmann}, \citenamefont {Kawasaki}, \citenamefont {Kohri},\ and\
  \citenamefont {Moroi}}]{Holtmann:1998gd}%
  \BibitemOpen
  \bibfield  {author} {\bibinfo {author} {\bibfnamefont {E.}~\bibnamefont
  {Holtmann}}, \bibinfo {author} {\bibfnamefont {M.}~\bibnamefont {Kawasaki}},
  \bibinfo {author} {\bibfnamefont {K.}~\bibnamefont {Kohri}}, \ and\ \bibinfo
  {author} {\bibfnamefont {T.}~\bibnamefont {Moroi}},\ }\href {\doibase
  10.1103/PhysRevD.60.023506} {\bibfield  {journal} {\bibinfo  {journal} {Phys.
  Rev. D}\ }\textbf {\bibinfo {volume} {60}},\ \bibinfo {pages} {023506}
  (\bibinfo {year} {1999})},\ \Eprint {http://arxiv.org/abs/hep-ph/9805405}
  {arXiv:hep-ph/9805405} \BibitemShut {NoStop}%
\bibitem [{\citenamefont {Kawasaki}\ \emph
  {et~al.}(2005{\natexlab{a}})\citenamefont {Kawasaki}, \citenamefont {Kohri},\
  and\ \citenamefont {Moroi}}]{Kawasaki:2004yh}%
  \BibitemOpen
  \bibfield  {author} {\bibinfo {author} {\bibfnamefont {M.}~\bibnamefont
  {Kawasaki}}, \bibinfo {author} {\bibfnamefont {K.}~\bibnamefont {Kohri}}, \
  and\ \bibinfo {author} {\bibfnamefont {T.}~\bibnamefont {Moroi}},\ }\href
  {\doibase 10.1016/j.physletb.2005.08.045} {\bibfield  {journal} {\bibinfo
  {journal} {Phys. Lett. B}\ }\textbf {\bibinfo {volume} {625}},\ \bibinfo
  {pages} {7} (\bibinfo {year} {2005}{\natexlab{a}})},\ \Eprint
  {http://arxiv.org/abs/astro-ph/0402490} {arXiv:astro-ph/0402490} \BibitemShut
  {NoStop}%
\bibitem [{\citenamefont {Kawasaki}\ \emph
  {et~al.}(2005{\natexlab{b}})\citenamefont {Kawasaki}, \citenamefont {Kohri},\
  and\ \citenamefont {Moroi}}]{Kawasaki:2004qu}%
  \BibitemOpen
  \bibfield  {author} {\bibinfo {author} {\bibfnamefont {M.}~\bibnamefont
  {Kawasaki}}, \bibinfo {author} {\bibfnamefont {K.}~\bibnamefont {Kohri}}, \
  and\ \bibinfo {author} {\bibfnamefont {T.}~\bibnamefont {Moroi}},\ }\href
  {\doibase 10.1103/PhysRevD.71.083502} {\bibfield  {journal} {\bibinfo
  {journal} {Phys. Rev. D}\ }\textbf {\bibinfo {volume} {71}},\ \bibinfo
  {pages} {083502} (\bibinfo {year} {2005}{\natexlab{b}})},\ \Eprint
  {http://arxiv.org/abs/astro-ph/0408426} {arXiv:astro-ph/0408426} \BibitemShut
  {NoStop}%
\bibitem [{\citenamefont {Kohri}\ and\ \citenamefont
  {Takayama}(2007)}]{Kohri:2006cn}%
  \BibitemOpen
  \bibfield  {author} {\bibinfo {author} {\bibfnamefont {K.}~\bibnamefont
  {Kohri}}\ and\ \bibinfo {author} {\bibfnamefont {F.}~\bibnamefont
  {Takayama}},\ }\href {\doibase 10.1103/PhysRevD.76.063507} {\bibfield
  {journal} {\bibinfo  {journal} {Phys. Rev. D}\ }\textbf {\bibinfo {volume}
  {76}},\ \bibinfo {pages} {063507} (\bibinfo {year} {2007})},\ \Eprint
  {http://arxiv.org/abs/hep-ph/0605243} {arXiv:hep-ph/0605243} \BibitemShut
  {NoStop}%
\bibitem [{\citenamefont {Kawasaki}\ \emph {et~al.}(2007)\citenamefont
  {Kawasaki}, \citenamefont {Kohri},\ and\ \citenamefont
  {Moroi}}]{Kawasaki:2007xb}%
  \BibitemOpen
  \bibfield  {author} {\bibinfo {author} {\bibfnamefont {M.}~\bibnamefont
  {Kawasaki}}, \bibinfo {author} {\bibfnamefont {K.}~\bibnamefont {Kohri}}, \
  and\ \bibinfo {author} {\bibfnamefont {T.}~\bibnamefont {Moroi}},\ }\href
  {\doibase 10.1016/j.physletb.2007.03.063} {\bibfield  {journal} {\bibinfo
  {journal} {Phys. Lett. B}\ }\textbf {\bibinfo {volume} {649}},\ \bibinfo
  {pages} {436} (\bibinfo {year} {2007})},\ \Eprint
  {http://arxiv.org/abs/hep-ph/0703122} {arXiv:hep-ph/0703122} \BibitemShut
  {NoStop}%
\bibitem [{\citenamefont {Kawasaki}\ \emph {et~al.}(2008)\citenamefont
  {Kawasaki}, \citenamefont {Kohri}, \citenamefont {Moroi},\ and\ \citenamefont
  {Yotsuyanagi}}]{Kawasaki:2008qe}%
  \BibitemOpen
  \bibfield  {author} {\bibinfo {author} {\bibfnamefont {M.}~\bibnamefont
  {Kawasaki}}, \bibinfo {author} {\bibfnamefont {K.}~\bibnamefont {Kohri}},
  \bibinfo {author} {\bibfnamefont {T.}~\bibnamefont {Moroi}}, \ and\ \bibinfo
  {author} {\bibfnamefont {A.}~\bibnamefont {Yotsuyanagi}},\ }\href {\doibase
  10.1103/PhysRevD.78.065011} {\bibfield  {journal} {\bibinfo  {journal} {Phys.
  Rev. D}\ }\textbf {\bibinfo {volume} {78}},\ \bibinfo {pages} {065011}
  (\bibinfo {year} {2008})},\ \Eprint {http://arxiv.org/abs/0804.3745}
  {arXiv:0804.3745 [hep-ph]} \BibitemShut {NoStop}%
\bibitem [{\citenamefont {Jittoh}\ \emph {et~al.}(2011)\citenamefont {Jittoh},
  \citenamefont {Kohri}, \citenamefont {Koike}, \citenamefont {Sato},
  \citenamefont {Sugai}, \citenamefont {Yamanaka},\ and\ \citenamefont
  {Yazaki}}]{Jittoh:2011ni}%
  \BibitemOpen
  \bibfield  {author} {\bibinfo {author} {\bibfnamefont {T.}~\bibnamefont
  {Jittoh}}, \bibinfo {author} {\bibfnamefont {K.}~\bibnamefont {Kohri}},
  \bibinfo {author} {\bibfnamefont {M.}~\bibnamefont {Koike}}, \bibinfo
  {author} {\bibfnamefont {J.}~\bibnamefont {Sato}}, \bibinfo {author}
  {\bibfnamefont {K.}~\bibnamefont {Sugai}}, \bibinfo {author} {\bibfnamefont
  {M.}~\bibnamefont {Yamanaka}}, \ and\ \bibinfo {author} {\bibfnamefont
  {K.}~\bibnamefont {Yazaki}},\ }\href {\doibase 10.1103/PhysRevD.84.035008}
  {\bibfield  {journal} {\bibinfo  {journal} {Phys. Rev. D}\ }\textbf {\bibinfo
  {volume} {84}},\ \bibinfo {pages} {035008} (\bibinfo {year} {2011})},\
  \Eprint {http://arxiv.org/abs/1105.1431} {arXiv:1105.1431 [hep-ph]}
  \BibitemShut {NoStop}%
\bibitem [{\citenamefont {Feng}\ \emph
  {et~al.}(2003{\natexlab{b}})\citenamefont {Feng}, \citenamefont {Rajaraman},\
  and\ \citenamefont {Takayama}}]{Feng:2003uy}%
  \BibitemOpen
  \bibfield  {author} {\bibinfo {author} {\bibfnamefont {J.~L.}\ \bibnamefont
  {Feng}}, \bibinfo {author} {\bibfnamefont {A.}~\bibnamefont {Rajaraman}}, \
  and\ \bibinfo {author} {\bibfnamefont {F.}~\bibnamefont {Takayama}},\ }\href
  {\doibase 10.1103/PhysRevD.68.063504} {\bibfield  {journal} {\bibinfo
  {journal} {Phys. Rev. D}\ }\textbf {\bibinfo {volume} {68}},\ \bibinfo
  {pages} {063504} (\bibinfo {year} {2003}{\natexlab{b}})},\ \Eprint
  {http://arxiv.org/abs/hep-ph/0306024} {arXiv:hep-ph/0306024} \BibitemShut
  {NoStop}%
\bibitem [{\citenamefont {Zhao}\ \emph {et~al.}(2013)\citenamefont {Zhao} \emph
  {et~al.}}]{Zhao:2012xw}%
  \BibitemOpen
  \bibfield  {author} {\bibinfo {author} {\bibfnamefont {G.-B.}\ \bibnamefont
  {Zhao}} \emph {et~al.},\ }\href {\doibase 10.1093/mnras/stt1710} {\bibfield
  {journal} {\bibinfo  {journal} {Mon. Not. Roy. Astron. Soc.}\ }\textbf
  {\bibinfo {volume} {436}},\ \bibinfo {pages} {2038} (\bibinfo {year}
  {2013})},\ \Eprint {http://arxiv.org/abs/1211.3741} {arXiv:1211.3741
  [astro-ph.CO]} \BibitemShut {NoStop}%
\bibitem [{\citenamefont {Palanque-Delabrouille}\ \emph
  {et~al.}(2013)\citenamefont {Palanque-Delabrouille} \emph
  {et~al.}}]{Palanque-Delabrouille:2013gaa}%
  \BibitemOpen
  \bibfield  {author} {\bibinfo {author} {\bibfnamefont {N.}~\bibnamefont
  {Palanque-Delabrouille}} \emph {et~al.},\ }\href {\doibase
  10.1051/0004-6361/201322130} {\bibfield  {journal} {\bibinfo  {journal}
  {Astron. Astrophys.}\ }\textbf {\bibinfo {volume} {559}},\ \bibinfo {pages}
  {A85} (\bibinfo {year} {2013})},\ \Eprint {http://arxiv.org/abs/1306.5896}
  {arXiv:1306.5896 [astro-ph.CO]} \BibitemShut {NoStop}%
\bibitem [{\citenamefont {Feng}\ \emph {et~al.}(2004)\citenamefont {Feng},
  \citenamefont {Su},\ and\ \citenamefont {Takayama}}]{Feng:2004zu}%
  \BibitemOpen
  \bibfield  {author} {\bibinfo {author} {\bibfnamefont {J.~L.}\ \bibnamefont
  {Feng}}, \bibinfo {author} {\bibfnamefont {S.-f.}\ \bibnamefont {Su}}, \ and\
  \bibinfo {author} {\bibfnamefont {F.}~\bibnamefont {Takayama}},\ }\href
  {\doibase 10.1103/PhysRevD.70.063514} {\bibfield  {journal} {\bibinfo
  {journal} {Phys. Rev. D}\ }\textbf {\bibinfo {volume} {70}},\ \bibinfo
  {pages} {063514} (\bibinfo {year} {2004})},\ \Eprint
  {http://arxiv.org/abs/hep-ph/0404198} {arXiv:hep-ph/0404198} \BibitemShut
  {NoStop}%
\bibitem [{\citenamefont {Kelso}\ \emph {et~al.}(2014)\citenamefont {Kelso},
  \citenamefont {de~S.~Pires}, \citenamefont {Profumo}, \citenamefont
  {Queiroz},\ and\ \citenamefont {Rodrigues~da Silva}}]{Kelso:2013nwa}%
  \BibitemOpen
  \bibfield  {author} {\bibinfo {author} {\bibfnamefont {C.}~\bibnamefont
  {Kelso}}, \bibinfo {author} {\bibfnamefont {C.~A.}\ \bibnamefont
  {de~S.~Pires}}, \bibinfo {author} {\bibfnamefont {S.}~\bibnamefont
  {Profumo}}, \bibinfo {author} {\bibfnamefont {F.~S.}\ \bibnamefont
  {Queiroz}}, \ and\ \bibinfo {author} {\bibfnamefont {P.~S.}\ \bibnamefont
  {Rodrigues~da Silva}},\ }\href {\doibase 10.1140/epjc/s10052-014-2797-3}
  {\bibfield  {journal} {\bibinfo  {journal} {Eur. Phys. J. C}\ }\textbf
  {\bibinfo {volume} {74}},\ \bibinfo {pages} {2797} (\bibinfo {year}
  {2014})},\ \Eprint {http://arxiv.org/abs/1308.6630} {arXiv:1308.6630
  [hep-ph]} \BibitemShut {NoStop}%
\bibitem [{\citenamefont {Rezaei}\ \emph {et~al.}(2019)\citenamefont {Rezaei},
  \citenamefont {Malekjani},\ and\ \citenamefont {Sola}}]{Rezaei:2019xwo}%
  \BibitemOpen
  \bibfield  {author} {\bibinfo {author} {\bibfnamefont {M.}~\bibnamefont
  {Rezaei}}, \bibinfo {author} {\bibfnamefont {M.}~\bibnamefont {Malekjani}}, \
  and\ \bibinfo {author} {\bibfnamefont {J.}~\bibnamefont {Sola}},\ }\href
  {\doibase 10.1103/PhysRevD.100.023539} {\bibfield  {journal} {\bibinfo
  {journal} {Phys. Rev. D}\ }\textbf {\bibinfo {volume} {100}},\ \bibinfo
  {pages} {023539} (\bibinfo {year} {2019})},\ \Eprint
  {http://arxiv.org/abs/1905.00100} {arXiv:1905.00100 [gr-qc]} \BibitemShut
  {NoStop}%
\bibitem [{\citenamefont {Rezaei}\ and\ \citenamefont
  {Sola~Peracaula}(2022)}]{Rezaei:2022bkb}%
  \BibitemOpen
  \bibfield  {author} {\bibinfo {author} {\bibfnamefont {M.}~\bibnamefont
  {Rezaei}}\ and\ \bibinfo {author} {\bibfnamefont {J.}~\bibnamefont
  {Sola~Peracaula}},\ }\href {\doibase 10.1140/epjc/s10052-022-10653-x}
  {\bibfield  {journal} {\bibinfo  {journal} {Eur. Phys. J. C}\ }\textbf
  {\bibinfo {volume} {82}},\ \bibinfo {pages} {765} (\bibinfo {year} {2022})},\
  \Eprint {http://arxiv.org/abs/2207.14250} {arXiv:2207.14250 [astro-ph.CO]}
  \BibitemShut {NoStop}%
\end{thebibliography}%

\end{document}